\documentclass[conference]{ieeeconf}  % Comment this line out
\IEEEoverridecommandlockouts                              % This command is only
\usepackage{graphics} % for pdf, bitmapped graphics files
\usepackage{epsfig} % for postscript graphics files
\usepackage{mathptmx} % assumes new font selection scheme installed
\usepackage{times} % assumes new font selection scheme installed
\usepackage{amsmath} % assumes amsmath package installed
\usepackage{amssymb}  % assumes amsmath package installed
\usepackage{color}
\usepackage{multirow}
\usepackage{footnote}
\usepackage{caption}
\usepackage[hyphens]{url} % Marc
\usepackage{hyperref}
\usepackage{float}
\usepackage{cuted}
\usepackage[table,xcdraw]{xcolor}
\usepackage{subcaption}
\usepackage{comment}
\usepackage{cite}
\usepackage{soul}

\title{\LARGE \bf
The TMB path loss model for 5 GHz indoor WiFi scenarios:\\ On the empirical relationship between RSSI, MCS, and spatial streams} 
\author{Toni Adame, Marc Carrascosa and Boris Bellalta}

\begin{document}

\maketitle
\thispagestyle{empty}
\pagestyle{empty}

%%%%%%%%%%%%%%%%%%%%%%%%%%%%%%%%%%%%%%%%%%%%%%%%%%%%%%%%%%%%%%%%%%%%%%%%%%%%%%%%
\begin{abstract}

The WiFi landscape is rapidly changing over the last years, responding to the new needs of wireless communications. IEEE 802.11ax is the next fast-approaching standard, addressing some of today’s biggest performance challenges specifically for high-density public environments. It is designed to operate at 2.4 GHz and 5 GHz bands, the latter being rapidly adopted worldwide  after its inclusion in IEEE 802.11ac, and with expected growing demand in the next 10 years. %Accurate sizing, planning and deployment will then become crucial in high-demanding scenarios for proper network operation.

This paper assesses empirically the suitability of the available IEEE 802.11ax path loss models at 5 GHz on some real testbeds and proposes a new model with higher abstraction level; i.e., without requiring from a previous \textit{in situ} analysis of each considered receiver's location. The proposed TMB path loss model, used in combination with generated data sets, is able to obtain an estimation of RSSI, selected modulation and coding scheme (MCS), and number of spatial streams in function of the AP configuration and the AP-STA distance. We aim to use the model to compare IEEE 802.11ac/ax performance simulation results with experimental ones.
\end{abstract}

%%%%%%%%%%%%%%%%%%%%%%%%%%%%%%%%%%%%%%%%%%%%%%%%%%%%%%%%%%%%%%%%%%%%%%%%%%%%%%%%
\section{Introduction}

%\textcolor{red}{Una crítica que ens poden fer és que no hi ha res d'estat de l'art, i.e., cap comentari d'altres models fets similars al nostre. Jo buscaria 3-4 papers que hagin fet més o menys el mateix, i el posaria a la intro com exemples de treballs similars, remarcant el que fem nosaltres de diferent (suposo que serà sobretot lo de MC, el data set, i la funció Matlab)}

IEEE 802.11ax is intended to replace both IEEE 802.11n and IEEE 802.11ac, targeting to improve the spectrum utilization efficiency, and working at both 2.4 GHz and 5 GHz frequency bands. The IEEE 802.11ax task group (TGax) \cite{tgax2018}, responsible for the design of the amendment named IEEE 802.11ax-2019, aims to improve PHY and MAC efficiency with modulation and coding schemes (MCSs) ranging from BPSK--1/2 to 1024-QAM--5/6.

TGax faces two main challenges: the ability of addressing dense scenarios and satisfying the increase of users' throughput needs \cite{bellalta2016ieee}. In fact, some of the main targeted use cases are indoors, such as crowded urban scenarios (apartment complexes, condominiums, and multi-dwelling buildings) or enterprise-class scenarios (next generation e-classrooms, colleges, and school campuses) \cite{qualcomm2018trans}. 

%Consequently, to meet this increasingly growing demand it is required a thorough access point (AP) deployment planning which considers the effects of indoor propagation.

To analyze the performance of this technology in such dense scenarios it is necessary to rely on simulators and analytical tools as realistic as possible. The use of these tools would then foster the design and development of advanced path loss / PHY models, statistical MAC protocols, as well as thorough access point (AP) deployment planning.

With this goal in mind, the current article evaluates the accuracy of the already available IEEE 802.11ax indoor path loss models at 5 GHz and compares it with the proposed empirical TMB model, which does not require from previous computation of traversed obstacles, unlike other similar models \cite{ lott2001multi, keenan1990radio, solahuddin2011indoor, xu2007indoor}. Besides, by combining it with other available data, the TMB model is able to provide the selected MCS and number of spatial streams for a given distance, in addition to the RSSI.

%\textcolor{red}{Aquí podriem posar un paràgraf dient que per estudiar les prestacions del 11ac i del 11ax es necessari tenir simulador i eines analítiques que puguin ser el més realistes possibles. Amb aquest objectiu en aquest paper validem si els models proposats per l'11ax TG funciona com s'espera, proposant una variació que permet no tenir en compte la posició de les estacions ni dels diferents 'obstacles', però que sobretot afegeix aspectes com proporcionar per cada punt el conjunt de transmission rates, i SS que es poden utilitzar.} 

%In this article, the already available IEEE 802.11ax indoor path loss models at 5 GHz have been evaluated by means of several experiments on different testbeds and compared with a proposed empirical model. The effect of indoor propagation on modulation and coding scheme (MCS) and spatial stream selection has also been studied, resulting in the \textit{MCS calculator} tool to accurately simulate the behavior of a network prior to its deployment. \textcolor{red}{[Jo parlaria de data set i d'un conjunt de funcions per treure info del data set. No parlaria en cap cas de la tool, ja que sembla que és el punt fort del paper, i és un script]}

The contributions of this paper can therefore be summarized into three main points: \begin{itemize}
    \item The validation of the IEEE 802.11ax path loss models in the 5 GHz band. %, including the effect on received signal of close by locations, adjacent channels, and long operational periods.
    \item The proposal of a more general path loss model that averages the effect of the different obstacles between transmitter and receiver, also providing information about the achievable data rates at a given distance (i.e., MCS and number of spatial streams).
    \item The generated data sets, including measurements from multiple locations in an indoor scenario, and the implemented MATLAB functions to extract the information of interest.
\end{itemize}

The remainder of this paper is organized as follows: Section \ref{path_loss} introduces IEEE 802.11ax indoor path loss models. Section \ref{testbeds} describes employed technology and considered testbeds. Next, Section \ref{experimentation} details the empirical process to obtain a new path loss model and compiles all obtained results from tests. Lastly, Section \ref{conclusions} presents the obtained conclusions and discusses open challenges.

%----------------------------------------------------------------------------------
%----------------------------------------------------------------------------------
%----------------------------------------------------------------------------------
%----------------------------------------------------------------------------------

\section{IEEE 802.11ax indoor path loss models}
\label{path_loss}

\begin{figure*}
\begin{eqnarray}
\text{PL}_{\text{res}}(d_{i,j}) = 40.05 + 20 \cdot \log_{10} \left(\frac{f_{c}}{2.4}\right) + 20 \cdot \log_{10} \left(\min \left\lbrace d_{i,j}, 5\right\rbrace \right) + (d_{i,j} > 5) \cdot 35 \cdot \log_{10} \left(\frac{d_{i,j}}{5}\right) + 18.3 \cdot F_{i,j}^{\frac{F_{i,j}+2}{F_{i,j}+1}-0.46} + 5 \cdot W_{i,j} 
\label{eq:residential}
\end{eqnarray}

\begin{eqnarray}
\text{PL}_{\text{ent}}(d_{i,j}) = 40.05 + 20 \cdot \log_{10} \left(\frac{f_{c}}{2.4}\right) + 20 \cdot \log_{10} \left(\min \left\lbrace d_{i,j}, 10\right\rbrace \right) + (d_{i,j} > 10) \cdot 35 \cdot \log_{10} \left(\frac{d_{i,j}}{10}\right) + 7 \cdot W_{i,j} 
\label{eq:enterprise}
\end{eqnarray}
\end{figure*}

IEEE 802.11ax adopts the IEEE 802.11ac channel model and penetration losses for link and system level performance evaluation in indoor scenarios \cite{tgax2018channel}. Specifically, IEEE 802.11ax standard defines 3 simulation scenarios \cite{tgax2018simulation, afaqui2016ieee}: 
\begin{enumerate}
\item \textbf{Residential:} In this environment, which models a 5-floor building with 20 apartments per floor, a large number of APs is installed in close vicinity, so that increased interference level can greatly affect devices performance within the network.
\item \textbf{Enterprise:} Similar to residential environment, enterprises are providing WiFi as their primary source of access to the Internet through a managed network. A large number of devices is considered in this office floor configuration, with 8 offices, 64 cubicles per office, and 4 stations per cubicle. 
\item \textbf{Indoor small basic service set (BSS):} This scenario captures the issues of representative real-world deployments with high density of APs and STAs, where the BSS from each operator is deployed in regular symmetry.\footnote{The indoor small BSS scenario has not been considered in the current study as it does not include the effect of typical surrounding walls in its corresponding path loss model.}
\end{enumerate}

Equations (\ref{eq:residential}) and (\ref{eq:enterprise}) obtained from \cite{tgax2018channel} describe the path loss model for the residential and the enterprise scenario, respectively. Table \ref{cons} compiles the main technical parameters from both scenarios.

\begin{table}[h]
\centering
\caption{IEEE 802.11ax path loss model parameters for residential and enterprise scenarios.}
\label{cons}
\begin{tabular}{|c|c|c|}
\hline
\textbf{Parameter} & \textbf{Description} & \textbf{Unit} \\ \hline
  $d_{i,j}$                 &        Distance to the AP            &        m       \\ \hline
            $f_{c}$       &                Frequency      &      GHz         \\ \hline
            $W_{i,j}$       & Number of traversed office walls  & walls               \\ \hline
            $F_{i,j}$       & Number of traversed floors                     & floors              \\ \hline
\end{tabular}
\end{table}

%----------------------------------------------------------------------------------
%----------------------------------------------------------------------------------
%----------------------------------------------------------------------------------
%----------------------------------------------------------------------------------

\section{Scenario overview and testbeds}
\label{testbeds}

The selected environment to validate the IEEE 802.11ax indoor path loss models at 5 GHz was the 2\textsuperscript{nd} floor, right wing of the Tanger building at Universitat Pompeu Fabra (UPF) facilities.\footnote{UPF Communication campus website:\\ \url{https://www.upf.edu/web/campus/tanger}.} This space is characterized by a 50 m long transversal corridor with office rooms at both sides from 20 $\text{m}^{2}$ to 32 $\text{m}^{2}$ (see Figure~\ref{fig:testbed123}).

Floors from offices and the corridor consist of ceramic tiles, while ceilings are made up of plaster. Space between offices is filled with plaster walls of 17 cm of thickness. As for doors and walls between offices and the main corridor, they have 8 cm of thickness and are made up of composite and plaster, respectively. The ceiling height of every room is 2.65 m.

Furniture within offices mainly consists of cabinets, tables, chairs, and drawers, all of them made up of composite or aluminum with some metallic elements. In addition, offices contain varied computer equipment such as screens, computer towers, and printers. 

Measurements were obtained during working hours with people performing their daily tasks (even occasionally walking along the corridor and in the rooms). Coexisting Internet wireless networks working at 2.4 GHz and 5 GHz were kept active.

\subsection{Hardware}
Due to the lack of available IEEE 802.11ax commercial hardware at the moment of writing this paper, tests were conducted on IEEE 802.11ac, as both standards operate at 5 GHz band and have equivalent channel models.
\begin{itemize}
\item AP TP-Link Archer C7 C1750 V4: This router supports IEEE 802.11ac standard delivering a combined wireless data transfer rate of up to 1.75 Gbps. Wireless speeds of up to 1300 Mbps over the 5 GHz band can be achieved.\footnote{TP-Link Archer C7 C1750 V4 datasheet: \url{https://static.tp-link.com/Archer\%20C7\%20Datasheet\%204.0.pdf}}

\begin{comment}
\begin{table}\centering
	\renewcommand*{\arraystretch}{1.25} % Height of the cell, gives enough space for a sqrt
	\setlength\tabcolsep{3pt}   % Space between text and lines
    \caption{IEEE 802.11ac rates in Mbps, being $Y_{c}$ the coding rate, and LGI and SGI the long and short guard interval, respectively. Data rate is doubled when using 2 spatial streams.}
   \label{rates}
		\begin{tabular}{|c|c|c|c|c|c|c|c|c|}
			\hline		
			\multirow{2}{*}{\textbf{MCS}} &  \multirow{2}{*}{\textbf{Mod.}}  &  \multirow{2}{*}{\textbf{ $Y_{c}$ }} & \multicolumn{2}{|c|}{ \textbf{20 MHz }} &  \multicolumn{2}{|c|}{\textbf{40 MHz} } &  \multicolumn{2}{|c|}{\textbf{80 MHz }}\\\cline{4-9}
			 &  &  & \textbf{LGI} & \textbf{SGI} & \textbf{LGI} & \textbf{SGI} & \textbf{LGI} & \textbf{SGI}\\\hline
		    	0 & BPSK & 1/2 & 6.5 & 7.2 & 13.5 & 15 & 29.3 & 32.5\\\hline
		    	1 & QPSK & 1/2 & 13	& 14.4 & 27 & 30 & 58.5 & 65\\\hline
			2 & QPSK & 3/4 & 19.5 & 21.7 & 40.5	& 45 & 87.8 & 97.5\\\hline
			3 & 16-QAM & 1/2 & 26 & 28.9 & 54 & 60 & 117 & 130\\\hline
			4 & 16-QAM & 3/4 & 39 & 43.3 & 81 & 90 & 175.5 & 195\\\hline
			5 & 64-QAM & 2/3 & 52 & 57.8 & 108 & 120 & 234 & 260\\\hline
			6 & 64-QAM & 3/4 & 58.5 & 65 & 121.5 & 135 & 263.3 & 292.5\\\hline
			7 & 64-QAM & 5/6 & 65 & 72.2 & 135 & 150 & 292.5 & 325\\\hline
			8 & 256-QAM & 3/4 & 78 & 86.7 & 162 & 180 & 351 & 390\\\hline
			9 & 256-QAM & 5/6 & N/A & N/A & 180 & 200 & 390 & 433.3\\\hline

		\end{tabular}	
	\end{table}
\end{comment}

\item Laptop Dell Latitude E5580: It is worth noting here that while the AP has three antennas, employed laptops have only two, thus having access to two spatial streams for up to 866.7 Mbps when using 80 MHz channels.\footnote{Dell Latitude 5580 owner's manual: \url{https://topics-cdn.dell.com/pdf/latitude-15-5580-laptop_owners-manual_en-us.pdf}}
\end{itemize}

\subsection{Software}

\begin{itemize}
\item Commercial firmware of the AP was replaced by the corresponding OpenWrt firmware.\footnote{OpenWrt Firmware for TP-Link Model Archer C7 AC1750: \url{https://openwrt.org/toh/hwdata/tp-link/tp-link_archer_c7_v4}}
\item Laptops ran Ubuntu 16.04 TLS with Linux Kernel 4.13.0-36.\footnote{Ubuntu Linux kernel image for version 4.13.0-36: \url{https://packages.ubuntu.com/xenial/linux-image-4.13.0-36-generic}}
\item iPerf 2.0.5 was the tool used in sender laptops to generate UDP traffic.\footnote{iPerf main website: \url{https://iperf.fr/}}
\item Wireshark 2.6.1 was the application employed for the capture and analysis of transmitted packets.\footnote{Wireshark main website: \url{https://www.wireshark.org/}}
\item Aircrack-ng 1.2 Beta 3 was used for packet capturing of raw IEEE 802.11 frames.\footnote{Aircrack-ng main website: \url{http://www.aircrack-ng.org/}}
\end{itemize}

\begin{table*}[]
\centering
\caption{Configuration summary of the different experiments.}
\label{test_config}
\begin{tabular}{|l|l|c|c|c|c|c|c|c|}
\hline
\multicolumn{2}{|l|}{\textbf{Experiment}}                                                                                    & \textbf{Testbed}                 & \textbf{\begin{tabular}[c]{@{}c@{}}Number of\\ positions\end{tabular}} & \textbf{\begin{tabular}[c]{@{}c@{}}Frequency\\ channel\end{tabular}} & \textbf{BW (MHz)} & \textbf{$P_{\text{TX}}$ (dBm)} & \textbf{\begin{tabular}[c]{@{}c@{}}TX duration\\ per repetition (s)\end{tabular}} & \textbf{\begin{tabular}[c]{@{}c@{}}Number\\ of samples \\ per repetition\end{tabular}} \\ \hline
\multirow{3}{*}{\textbf{\begin{tabular}[c]{@{}l@{}}A. Signal \\ variance\end{tabular}}} & \textit{\textbf{Time effect}}      & Testbed \#2                      & 3                                                                      & 36                                                                   & 20                & 23                 & 600                                                                               &    $\approx$ 50000                                                                                    \\ \cline{2-9} 
                                                                                        & \textit{\textbf{Space effect}}     & \multicolumn{1}{l|}{Testbed \#3} & 3 x 9                                                                  & 36                                                                   & 20                & 23                 & 10                                                                                &                               $\approx$ 850                                                           \\ \cline{2-9} 
                                                                                        & \textit{\textbf{Frequency effect}} & \multicolumn{1}{l|}{Testbed \#2} & 3                                                                      & 36, 40, 44                                                           & 20                & 23                 & 10                                                                                &                                  $\approx$ 1700                                                         \\ \hline
\multicolumn{2}{|l|}{\textbf{B. Path loss}}                                                                                  & Testbed \#1                      & 21                                                                     & 36                                                                   & 20, 40, 80        & 4, 10, 23          & 10                                                                                &                   $\approx$ 850                                                                     \\ \hline
\multicolumn{2}{|l|}{\textbf{\begin{tabular}[c]{@{}l@{}}C. Modulation and \\ coding scheme (MCS)\end{tabular}}}              & Testbed \#2                      & 3                                                                      & 36                                                                   & 20, 40, 80        & 4, 10, 23          & 10                                                                                &                           $\approx$ 850                                                                \\ \hline
\multicolumn{2}{|l|}{\textbf{D. Spatial streams}}                                                                            & Testbed \#1                      & 21                                                                     & 36                                                                   & 20, 40, 80        & 4, 10, 23          & 10                                                                                &                     $\approx$ 850                                                                      \\ \hline
\end{tabular}
\end{table*}
\begin{figure*}[h!]
\centering
\includegraphics[width=17cm]{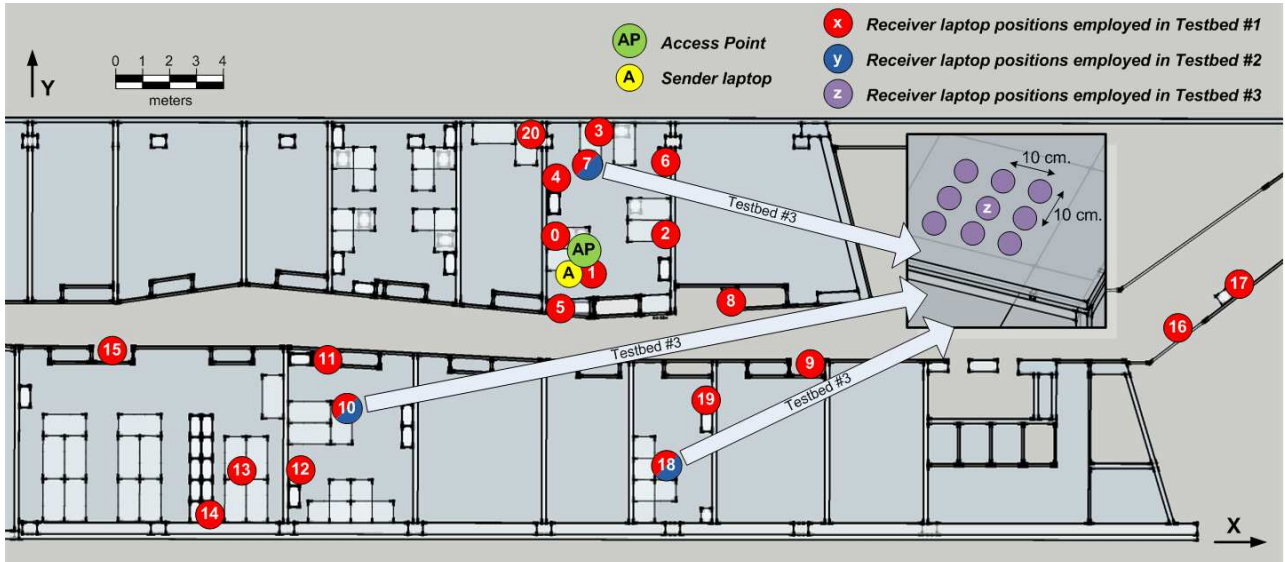}
\caption{Scenario floor plan and device deployment in Testbeds \#1, \#2, and \#3.}
\label{fig:testbed123}
\end{figure*}
\subsection{Testbeds}
\begin{itemize}

\item \textbf{Testbed \#1:} Full deployment in 21 locations

This testbed was made up of an AP and two laptops. The AP always maintained the same position on a table from a central office and was connected through an Ethernet cable to the sender laptop (laptop \textbf{\textit{A}}).

Measurements were taken in the receiver laptop, which occupied one of the $N_{L}$ = 21 pre-selected locations in every test repetition (see Table~\ref{locations} for position details and Figure~\ref{fig:testbed123} for device deployment), with the goal of covering a wide range of channel propagation cases.

\begin{table}[h!]
\centering
\caption{Summary of analyzed locations in Testbed \#1.}
\label{locations}
\begin{tabular}{|c|c|c|c|}
\hline
\textbf{Location} & \textbf{Height (m)} & \textbf{\begin{tabular}[c]{@{}c@{}}Distance to\\ the AP (m)\end{tabular}} &  \textbf{\begin{tabular}[c]{@{}c@{}}Traversed\\office walls\end{tabular}}
\\
\hline
\textbf{AP}       & 0.740               & -      &  -                                                                 \\ \hline
\textbf{0}        & 0.740               & 1.000   &  0                                                                \\ \hline
\textbf{1}        & 0.505               & 0.934  & 0                                                                  \\ \hline
\textbf{2}        & 0.740               & 3.084  & 0                                                                  \\ \hline
\textbf{3}        & 0.740               & 4.266  & 0                                                                  \\ \hline
\textbf{4}        & 1.680               & 2.717  & 0                                                                  \\ \hline
\textbf{5}        & 1.970               & 2.879  & 0                                                                  \\ \hline
\textbf{6}        & 1.680               & 3.995  & 0                                                                  \\ \hline
\textbf{7}        & 0.740               & 2.945  & 0                                                                  \\ \hline
\textbf{8}        & 0.505               & 5.778  & 2                                                                  \\ \hline
\textbf{9}        & 1.800               & 9.286  & 1                                                                  \\ \hline
\textbf{10}       & 0.740               & 11.141 & 4                                                                  \\ \hline
\textbf{11}       & 1.970               & 10.669 & 3                                                                  \\ \hline
\textbf{12}       & 1.970               & 13.884 & 4                                                                  \\ \hline
\textbf{13}       & 0.740               & 15.801 & 4                                                                  \\ \hline
\textbf{14}       & 1.970               & 17.579 & 5                                                                 \\ \hline
\textbf{15}       & 1.800               & 18.508 & 3                                                                  \\ \hline
\textbf{16}       & 0                   & 22.020 & 2                                                                  \\ \hline
\textbf{17}       & 0.505               & 24.304 & 2                                                                  \\ \hline
\textbf{18}       & 0.740               & 8.975 & 3                                                                   \\ \hline
\textbf{19}       & 1.970               & 7.267 & 2                                                                   \\ \hline
\textbf{20}       & 0.740               & 4.623 & 1                                                                   \\ \hline
\end{tabular}
\end{table}

\item \textbf{Testbed \#2:} Subset of locations

This testbed maintained the same locations for the AP and the sender laptop (laptop \textbf{\textit{A}}) from Testbed \#1, and reused locations \#7, \#10, and \#18 for the receiver laptop (see Figure~\ref{fig:testbed123} for device deployment). Again, measurements were taken in the receiver laptop.

\item \textbf{Testbed \#3:} Subset of STAs and close locations

Again, this testbed maintained the same position for the AP and the sender laptop (laptop \textbf{\textit{A}}) from Testbeds \#1 and \#2, and reused locations from Testbed \#2 for the receiver laptop. In addition, a 3x3 grid with a separation of 10 cm was created around each of the 3 selected locations, as shown in Figure~\ref{fig:testbed123}.
\end{itemize}

\subsection{Extraction of data sets}

Once deployed a testbed,  measurements were obtained as follows: firstly, a bandwidth-transmission power level (BW-$P_{\text{TX}}$) combination was set in the AP by means of the OpenWrt firmware. Then, the sender laptop (which was connected through an Ethernet cable to the AP) used iPerf to inject a determined UDP traffic load at 1 Mbps. Lastly, the receiver laptop running Wireshark captured several metrics from each received packet and stored them into \textit{.txt} files.

\section{Experimentation and results}
\label{experimentation}

Four different experiments were defined to study the behaviour of different IEEE 802.11ax parameters on the aforementioned testbeds: signal variance, path loss, MCS, and spatial streams. Table~\ref{test_config} compiles the main features of experiments, which are fully described in the following lines.

\subsection{Signal variance}
Prior to the in-depth analysis of path loss, effects of time, space, and frequency on received signal were studied. This quantification and assessment of signal variance was aimed to validate the procedure followed to get the measurements that would be used to obtain the TMB path loss model.
 \begin{figure*}[h!]
    \centering
    \includegraphics[width=13cm]{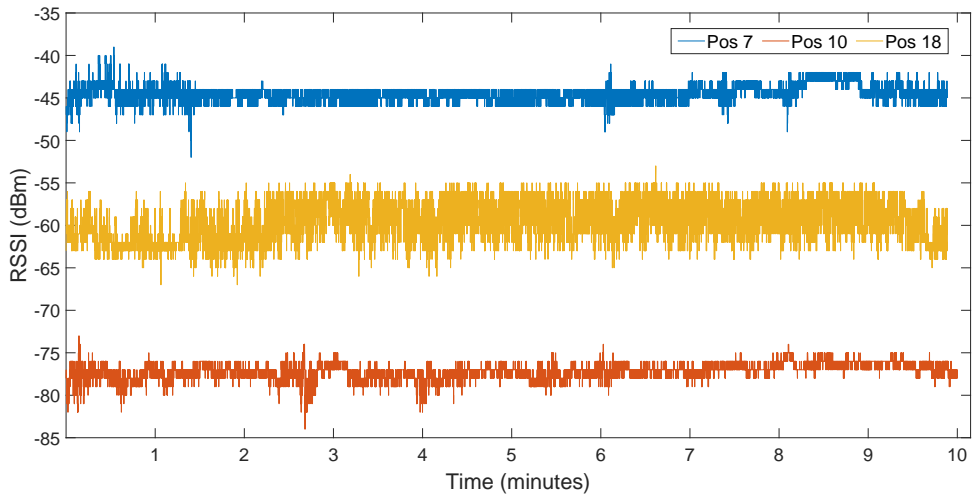}
    \caption{RSSI over time in Testbed \#2.}
    \label{fig:10min}
    \end{figure*}

\begin{itemize}
    \item \textit{Time effect} on signal variance was studied by sending a 10-minute continuous data stream to receiver locations from Testbed \#2. The full set of collected received signal strength indicator (RSSI) values over time is shown in Figure~\ref{fig:10min}. Except from some signal variance in the first minutes, channel maintained its stability for the whole test duration in all considered locations. Besides, computed standard deviation kept below the $\sigma =$ 5 dB of inherent shadowing defined in the IEEE 802.11ax simulation scenarios \cite{tgax2018simulation}, with values of 0.92 dB, 0.94 dB and 2.26 dB in locations \#7, \#10 and \#18, respectively.

    \begin{figure*}[h]
\centering
\begin{subfigure}[b]{0.32\textwidth}
                \includegraphics[width=\textwidth]{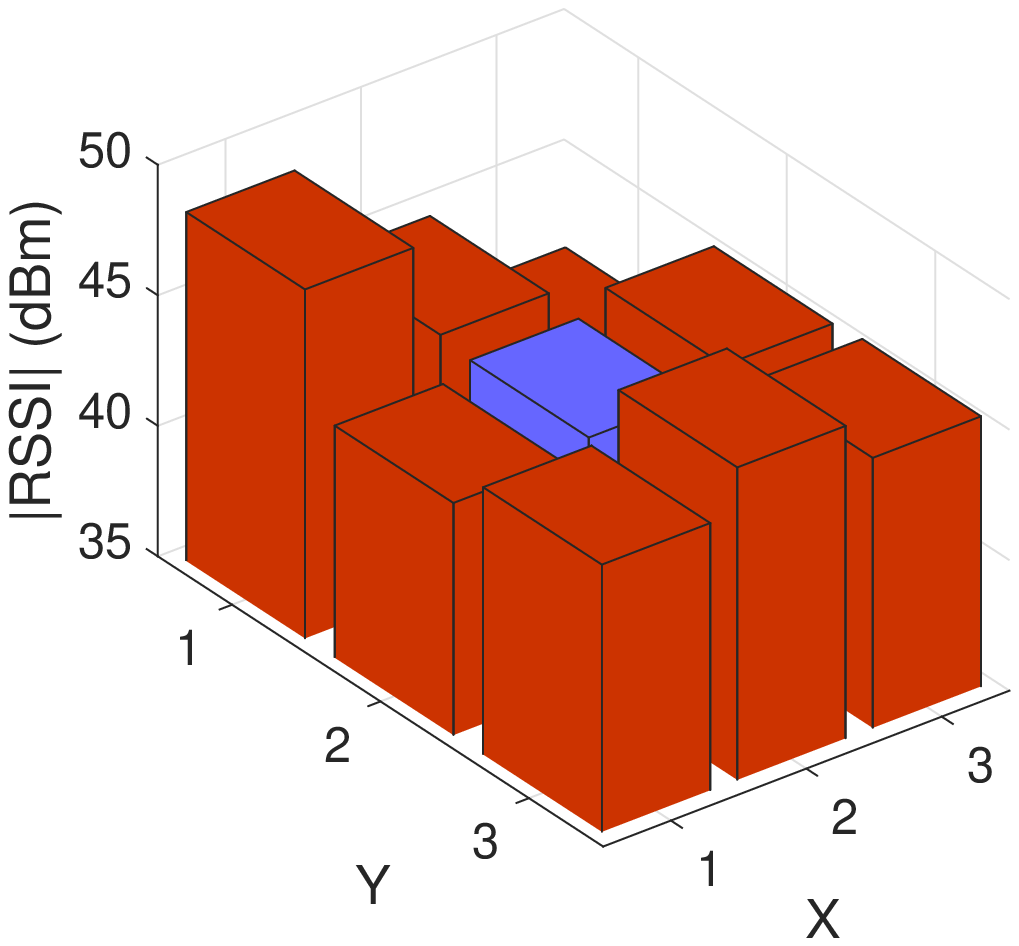}
                \caption{Location \#7.}
                \label{fig:grid_pos7}
        \end{subfigure}
     \begin{subfigure}[b]{0.32\textwidth}
                \includegraphics[width=\textwidth]{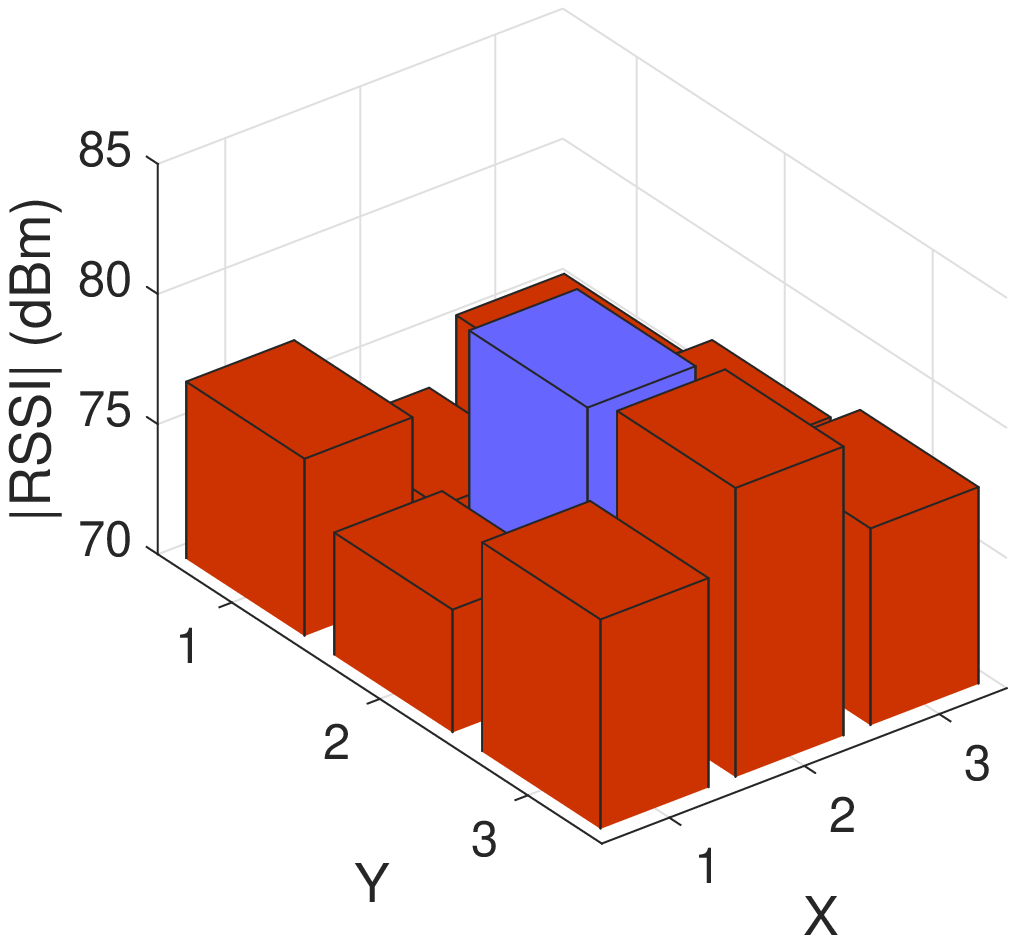}
                \caption{Location \#10.}
                \label{fig:fig:grid_pos10}
        \end{subfigure}   
      \begin{subfigure}[b]{0.32\textwidth}                \includegraphics[width=\textwidth]{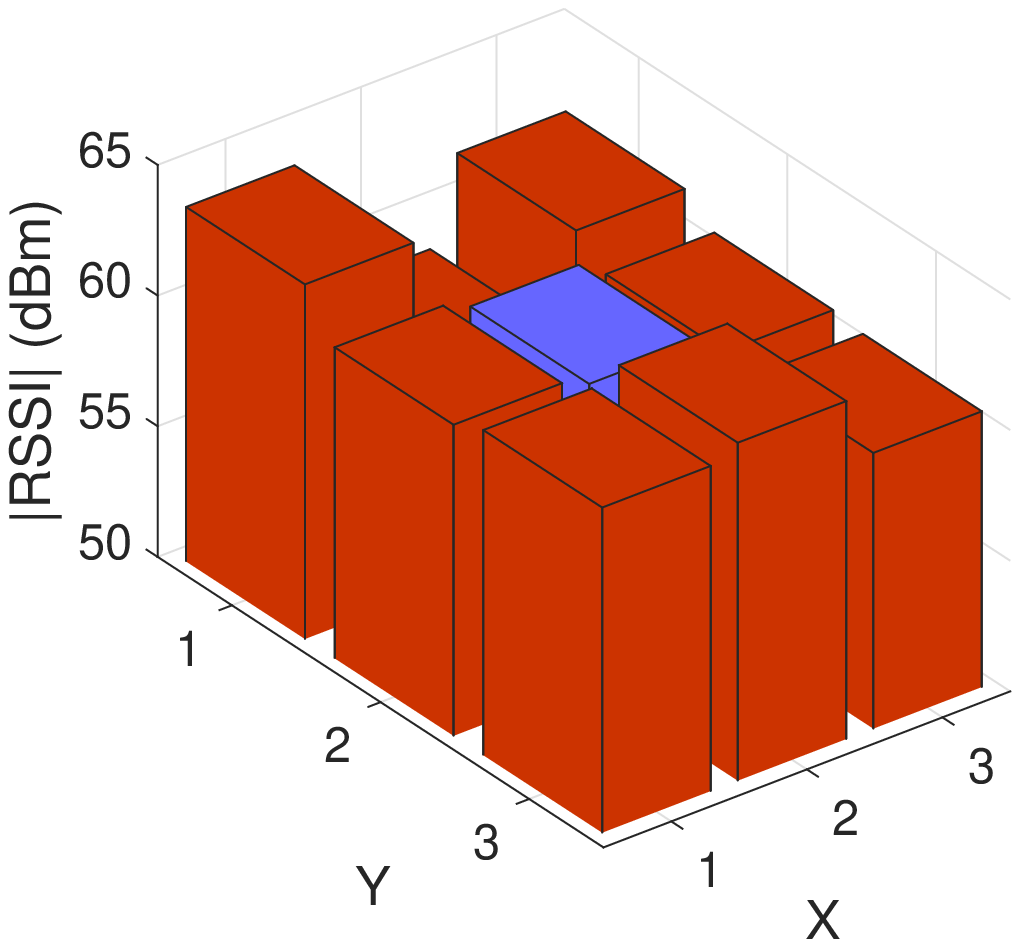}
                \caption{Location \#18.}
                \label{fig:fig:grid_pos18}
        \end{subfigure}
                \caption{Signal variance according to RSSI in 9-point grids deployed in Testbed \#3.}
        \label{fig:grids}
\end{figure*}
    \item \textit{Space effect} on signal variance was studied after injecting a 10-second continuous data stream per each 9-point grid of the 3 selected locations from Testbed \#3. Figure~\ref{fig:grids} shows the averaged RSSI value in every 9-point grid, where the reference value (i.e., the one in grid's central position) is always contained between the maximum and the minimum RSSI of its corresponding grid. More specifically, the maximum absolute difference observed with respect to the reference value is 3.96 dB, 3.38 dB and 3.11 dB in locations \#7, \#10 and \#18, respectively. As these values were below the $\sigma =$ 5 dB of inherent shadowing, it was assumed that any future measurement taken within a 20 cm x 20 cm squared area would correspond to the same location.
    
    \item Lastly, \textit{frequency effect} was analyzed by sending a 10-second continuous data stream to receiver locations from Testbed \#2 at three adjacent frequency channels: 36 ($f_{c}$~=~5.180 GHz), 40 ($f_{c}$~=~5.200 GHz), and 44 ($f_{c}$~=~5.220 GHz). Table~\ref{frequency} shows how RSSI differences with respect to the reference value from channel 36 were again confined below $\sigma =$ 5 dB, reflecting no significant influence of frequency channel on collected RSSI.\footnote{From that moment on, tests were conducted over channel 36.}
\end{itemize}

\begin{table}[]
\centering
\caption{Averaged RSSI and difference with the reference value in the frequency study performed in Testbed \#2.}
\label{frequency}
\begin{tabular}{|l|c|c|c|}
\hline
                                                                          & \textbf{Location \#7}                                           & \textbf{Location \#10}                                          & \textbf{Location \#18}                                          \\ \hline
\textbf{\begin{tabular}[c]{@{}l@{}}Channel 36\\ (reference)\end{tabular}} & -44.74 dBm                                                      & -78.43 dBm                                                      & -58.81 dBm                                                      \\ \hline
\textbf{Channel 40}                                                       & \begin{tabular}[c]{@{}c@{}}-43.40 dBm\\ (+1.34 dB)\end{tabular} & \begin{tabular}[c]{@{}c@{}}-77.14 dBm\\ (+1.29 dB)\end{tabular} & \begin{tabular}[c]{@{}c@{}}-57.91 dBm\\ (+0.90 dB)\end{tabular} \\ \hline
\textbf{Channel 44}                                                       & \begin{tabular}[c]{@{}c@{}}-40.96 dBm\\ (+3.78 dB)\end{tabular} & \begin{tabular}[c]{@{}c@{}}-77.84 dBm\\ (+0.59 dB)\end{tabular} & \begin{tabular}[c]{@{}c@{}}-59.56 dBm\\ (-0.75 dB)\end{tabular} \\ \hline
\end{tabular}
\end{table}

\subsection{Path loss}

With the goal of quantifying the channel propagation losses at 5 GHz and assessing the suitability of the residential and enterprise channel models proposed in the IEEE 802.11ax standard, a comprehensive study on the path loss was performed.

In each of the $N_{L}$ = 21 considered locations from Testbed \#1 in which the receiver laptop was placed, the sender laptop continuously sent it data packets through the AP for 10 seconds at a rate of 1 Mbps. This operation was repeated 9 times, one per each possible BW-$P_{\text{TX}}$ combination from Table~\ref{test_config}.\footnote{Available RSSI values from different bandwidths (BWs) were averaged and grouped into a single model, as no significant differences were found except from edge cases (i.e., close to the sensitivity level of the receiver), where higher BW values were slightly prone to signal loss.} RSSI values at the receiver laptop were used to compute the corresponding path loss (PL) according to
\begin{eqnarray}
\text{PL} = P_{\text{TX}} - P_{\text{RX}} = P_{\text{TX}} - \text{RSSI},
\label{eq:rssi}
\end{eqnarray}
where $P_{\text{TX}}$ and $P_{\text{RX}}$ correspond to the transmitted and received power, respectively. 

\begin{figure*}[h!]
\centering
\includegraphics[width=18cm]{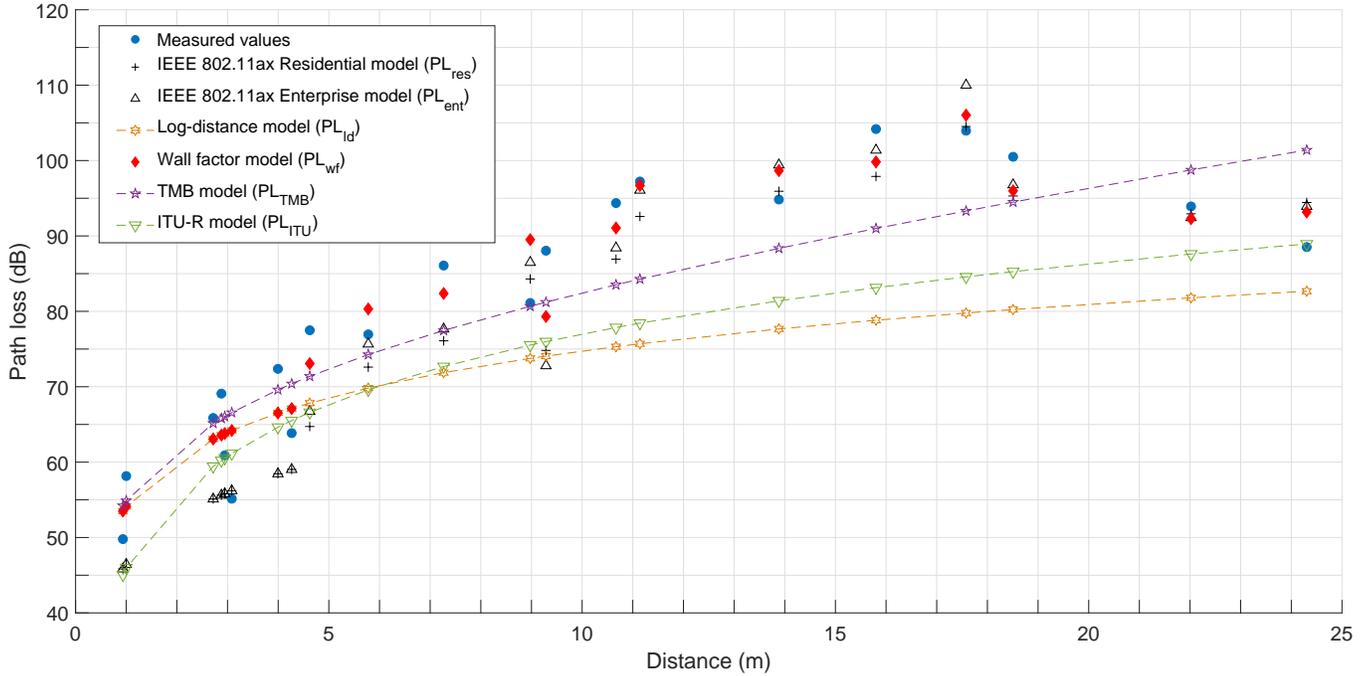}
\caption{Comparison of measured values and IEEE 802.11ax path loss models in Testbed \#1.}
\label{fig:PL_final}
\end{figure*}

Figure~\ref{fig:PL_final} shows path loss \textit{measured values} together with a representation of IEEE 802.11ax path loss models for residential and enterprise scenarios by means of (\ref{eq:residential}) and (\ref{eq:enterprise}), respectively. Similarity between both models and the actual measured values is noticeable even in the three furthest locations (\#15, \#16, and \#17), where the path loss is less than in closer locations, due to the low number of traversed office walls and the effect of the corridor. The $W_{i,j}$ factor, specific to the number of traversed walls for each location, makes the models suit to the current scenario. 

By taking into account the measured values in Testbed \#1, the following lines elaborate on the design of a generalizable path loss model. Firstly, the log-distance model
\begin{eqnarray}
\text{PL}_{\text{ld}}(d_{i,j}) = L_{0} + 10 \cdot \gamma \cdot \log_{10}(d_{i,j})
\label{eq:model_ld}
\end{eqnarray}
is obtained after applying a robust regression on path loss measured values from locations with no traversed walls (i.e., from \#0 to \#7), where $L_{0}$ is the path loss intercept and $\gamma$ is the attenuation factor.

Then, and as in \cite{xu2007indoor}, a new $k \cdot W_{i,j}$ factor is added to the previous model to define the wall factor path loss model
\begin{eqnarray}
\text{PL}_{\text{wf}}(d_{i,j}) = L_{0} + 10 \cdot \gamma \cdot \log_{10}(d_{i,j}) + k \cdot W_{i,j},
\label{eq:model_wf}
\end{eqnarray}
being $k$ the attenuation of each wall, and $W_{i,j}$ the number of traversed walls. $k$ is chosen as the value which minimizes the root mean square error (RMSE) with measured values.

Lastly, with the goal of avoiding the use of a \textit{location-specific} value like $W_{i,j}$, the TMB model  
\begin{eqnarray}
\text{PL}_{\text{TMB}}(d_{i,j}) = L_{0} + 10 \cdot \gamma \cdot \log_{10}(d_{i,j}) + k \cdot \overline{W} \cdot d_{i,j}
\label{eq:model_tmb}
\end{eqnarray}
is proposed. In this case, $W_{i,j}$ value is replaced by the \textit{distance-dependent} expression $W(d_{i,j}) = \overline{W} \cdot d_{i,j}$, where $\overline{W}$ is the average number of traversed walls per meter in the $N_{L}$ analyzed locations \cite{wilhelmi2018potential}.

Additionally, and for comparison purposes, the ITU-R indoor site-general model   
\begin{eqnarray}
\text{PL}_{\text{ITU}}(d_{i,j}) = 20 \cdot \log_{10}(f_{c}) + N \cdot \log_{10}(d_{i,j}) + L_{f} - 28,
\label{eq:model_itu}
\end{eqnarray}
is considered \cite{itu2012p}, where $f_{c}$ is the employed frequency, $N$ is the distance power loss coefficient (in our particular case and according to the model guidelines, $N = 31$), and $L_{f}$ is the floor penetration loss factor (which was removed as experimentation was performed on a single floor).

Figure~\ref{fig:PL_final} compares the aforementioned path loss models with measured values in Testbed \#1. While the ITU-R and specially the log-distance model are not able to reflect the effect of walls on indoor propagation, the wall factor model (due to the introduction of the $W_{i,j}$ value for each location) better matches with real behavior. The TMB model, for its part, also characterizes channel propagation better than the log-distance and the ITU-R models, thanks to the introduction of the averaged \textit{wall attenuation factor} $\overline{W}$. 

Table~\ref{parameters} compiles all parameters introduced in the aforementioned models as well as the obtained RMSE when comparing them to the path loss measured values from Testbed \#1. Low error is achieved in \textit{location-specific} models (i.e., residential, enterprise, and wall factor), specially in the wall factor one, but it is necessary to know the number of traversed walls for each receiver's location. 

As for the \textit{continuous} models (i.e., log-distance, TMB, and ITU-R), it is worth noting the behavior of the TMB model, which outperforms IEEE 802.11ax residential and enterprise models, thus proving its suitability in indoor scenarios.

\begin{table}[]
\centering
\caption{Main parameters and RMSE of path loss models analyzed in Testbed \#1.}
\label{parameters}
\begin{tabular}{|c|c|c|c|c|c|}
\hline
\multirow{2}{*}{\textbf{\begin{tabular}[c]{@{}c@{}}Path loss\\ model\end{tabular}}}                                    & \multicolumn{4}{c|}{\textbf{Parameters}}      & \multirow{2}{*}{\textbf{RMSE (dB)}} \\ \cline{2-5}
                                                          &  $L_{0}$     & $\gamma$    & $k$    & $\overline{W}$ &                            \\ \hline
\begin{tabular}[c]{@{}c@{}}Residential\\ $\text{PL}_{\text{res}}(d_{i,j})$\end{tabular} & \multicolumn{4}{c|}{\textit{see Equation (\ref{eq:residential})}} & 7.9932                   \\ \hline
\begin{tabular}[c]{@{}c@{}}Enterprise\\ $\text{PL}_{\text{ent}}(d_{i,j})$\end{tabular}  & \multicolumn{4}{c|}{\textit{see Equation (\ref{eq:enterprise})}} & 7.8431                  \\ \hline
\begin{tabular}[c]{@{}c@{}}Log-distance\\ $\text{PL}_{\text{ld}}(d_{i,j})$\end{tabular}  &  54.1200      & 2.06067         &  -    &  - &     13.3454                   \\ \hline
\begin{tabular}[c]{@{}c@{}}Wall Factor\\ $\text{PL}_{\text{wf}}(d_{i,j})$\end{tabular}     &  54.1200       & 2.06067        &  5.25    & -    & 4.8237                   \\ \hline
\begin{tabular}[c]{@{}c@{}}TMB\\ $\text{PL}_{\text{TMB}}(d_{i,j})$\end{tabular}    &  54.1200      &  2.06067        &  5.25    & 0.1467      &   7.7283           \\ \hline
\begin{tabular}[c]{@{}c@{}}ITU-R\\ $\text{PL}_{\text{ITU}}(d_{i,j})$\end{tabular}    & \multicolumn{4}{c|}{\textit{see Equation (\ref{eq:model_itu})}} &  11.5772          \\ \hline
\end{tabular}
\end{table}

\subsection{Modulation and coding scheme (MCS)}

The dynamic selection by the AP of the most appropriate MCS in function of the node location was analyzed in Testbed \#2. The receiver laptop was alternatively placed in locations \#7, \#10, and \#18, receiving a 10-second data flow from the AP at a rate of 1 Mbps. This operation was repeated 9 times, one per each possible BW-$P_{\text{TX}}$ combination from Table~\ref{test_config}. At every repetition, the MCS value of every packet was examined and stored.

\begin{figure*}[]
\centering
\begin{subfigure}[b]{0.32\textwidth}
                \includegraphics[width=\textwidth]{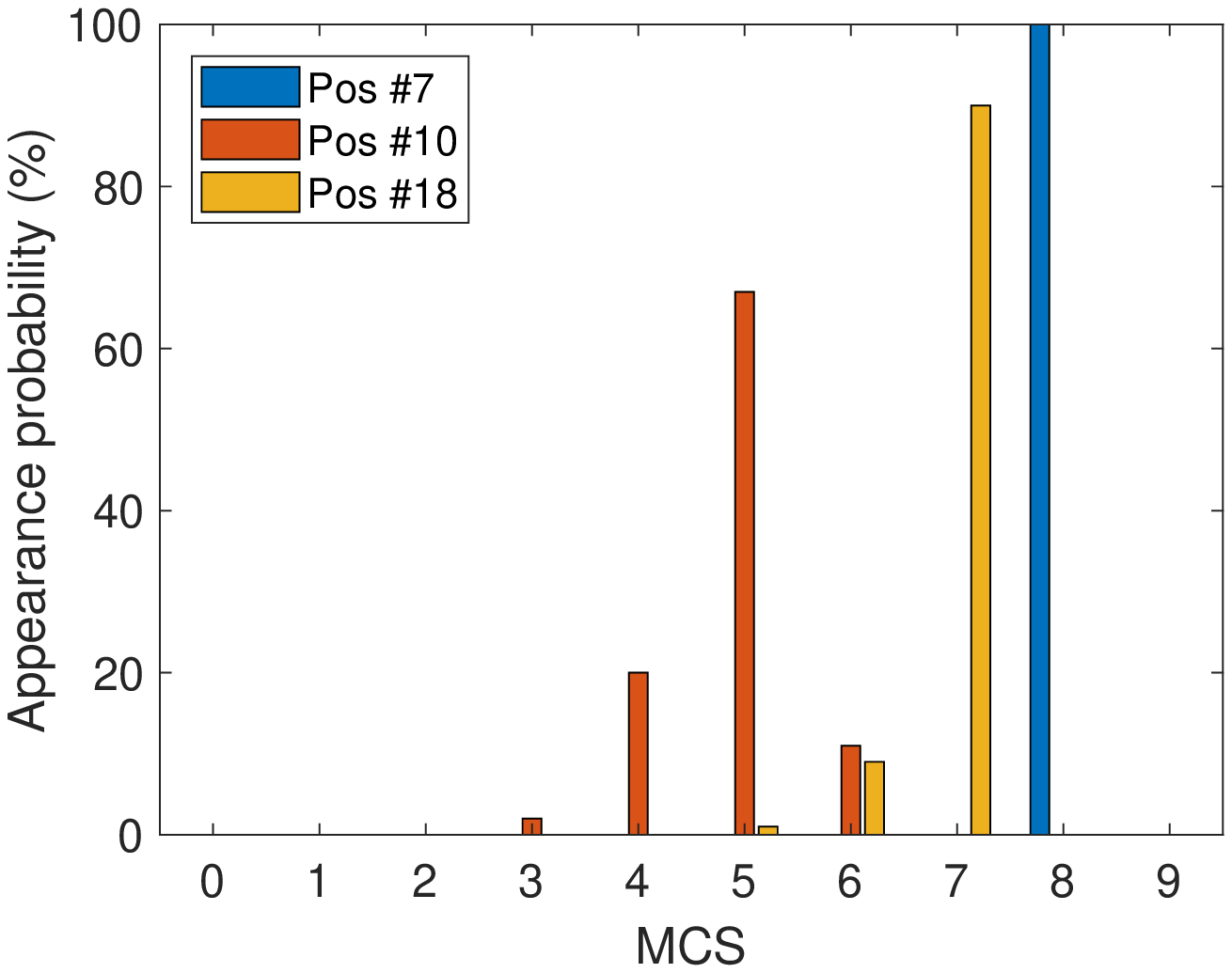}
                \caption{BW = 20 MHz, $P_{\text{TX}}$ = 23 dBm}
                \label{fig:MCS_20_23}
        \end{subfigure}
     \begin{subfigure}[b]{0.32\textwidth}
                \includegraphics[width=\textwidth]{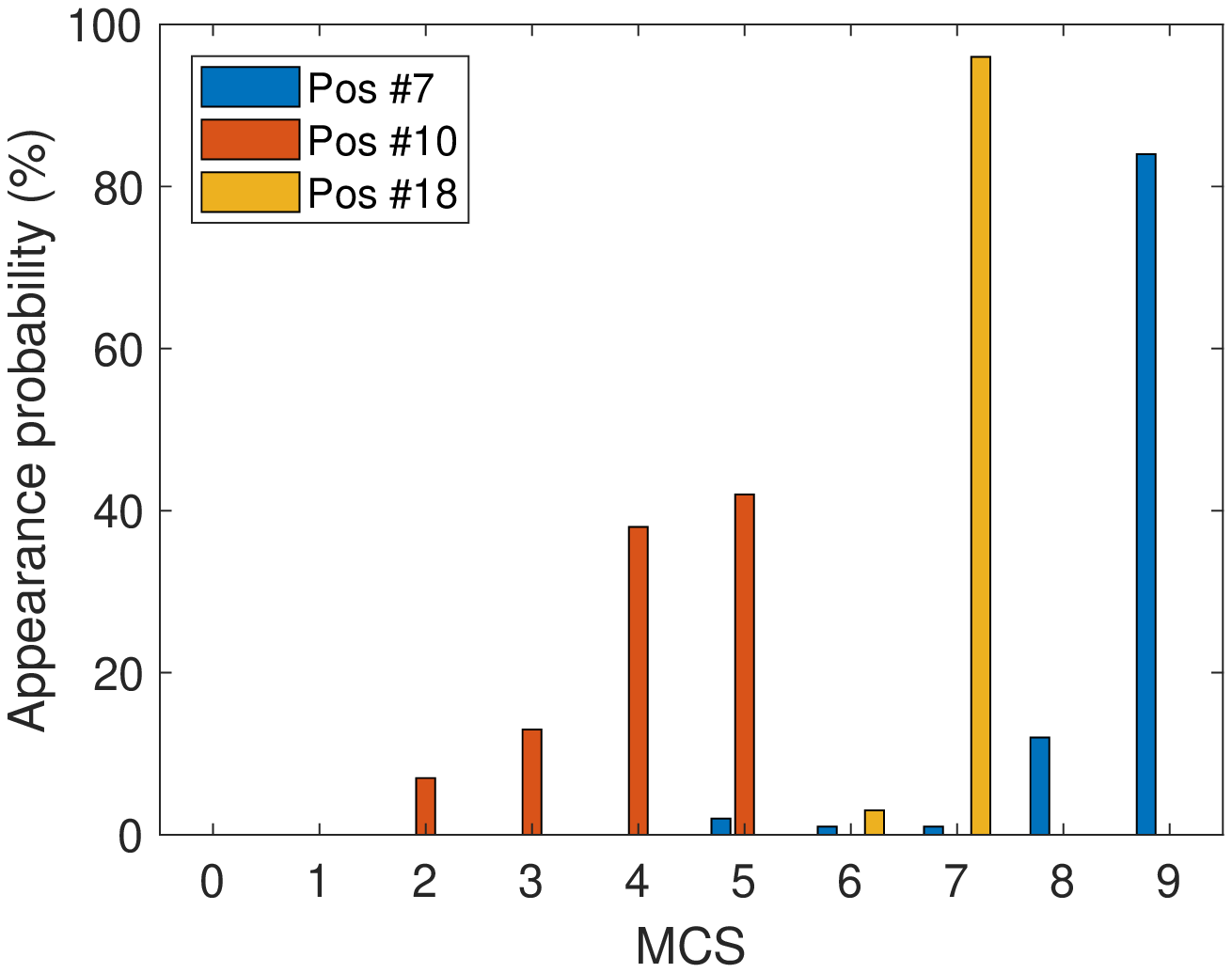}
                \caption{BW = 40 MHz, $P_{\text{TX}}$ = 23 dBm}
                \label{fig:MCS_40_23}
        \end{subfigure}   
      \begin{subfigure}[b]{0.32\textwidth}                \includegraphics[width=\textwidth]{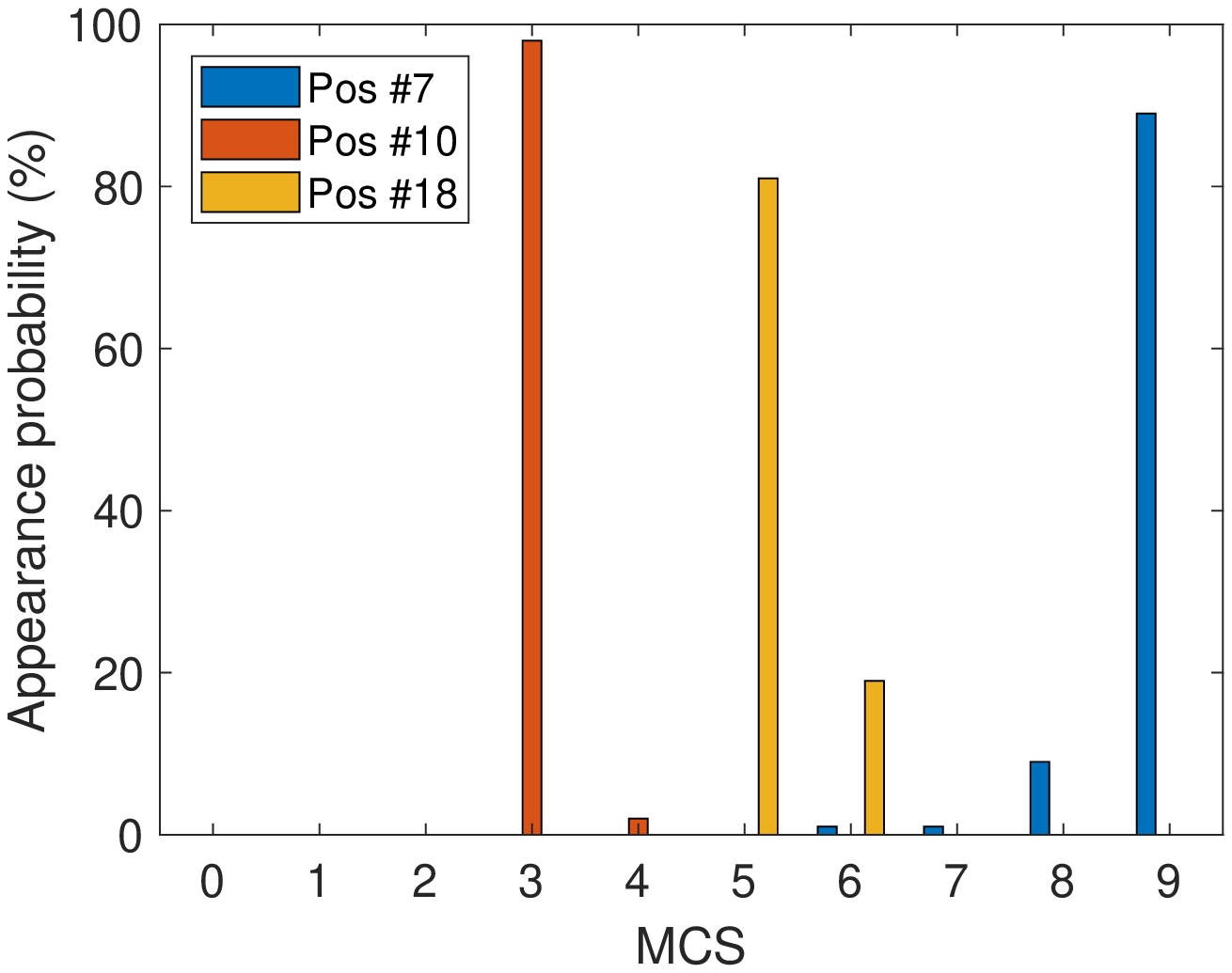}
                \caption{BW = 80 MHz, $P_{\text{TX}}$ = 23 dBm}
                \label{fig:MCS_80_23}
        \end{subfigure}
      
    \begin{subfigure}[b]{0.32\textwidth}
                \includegraphics[width=\textwidth]{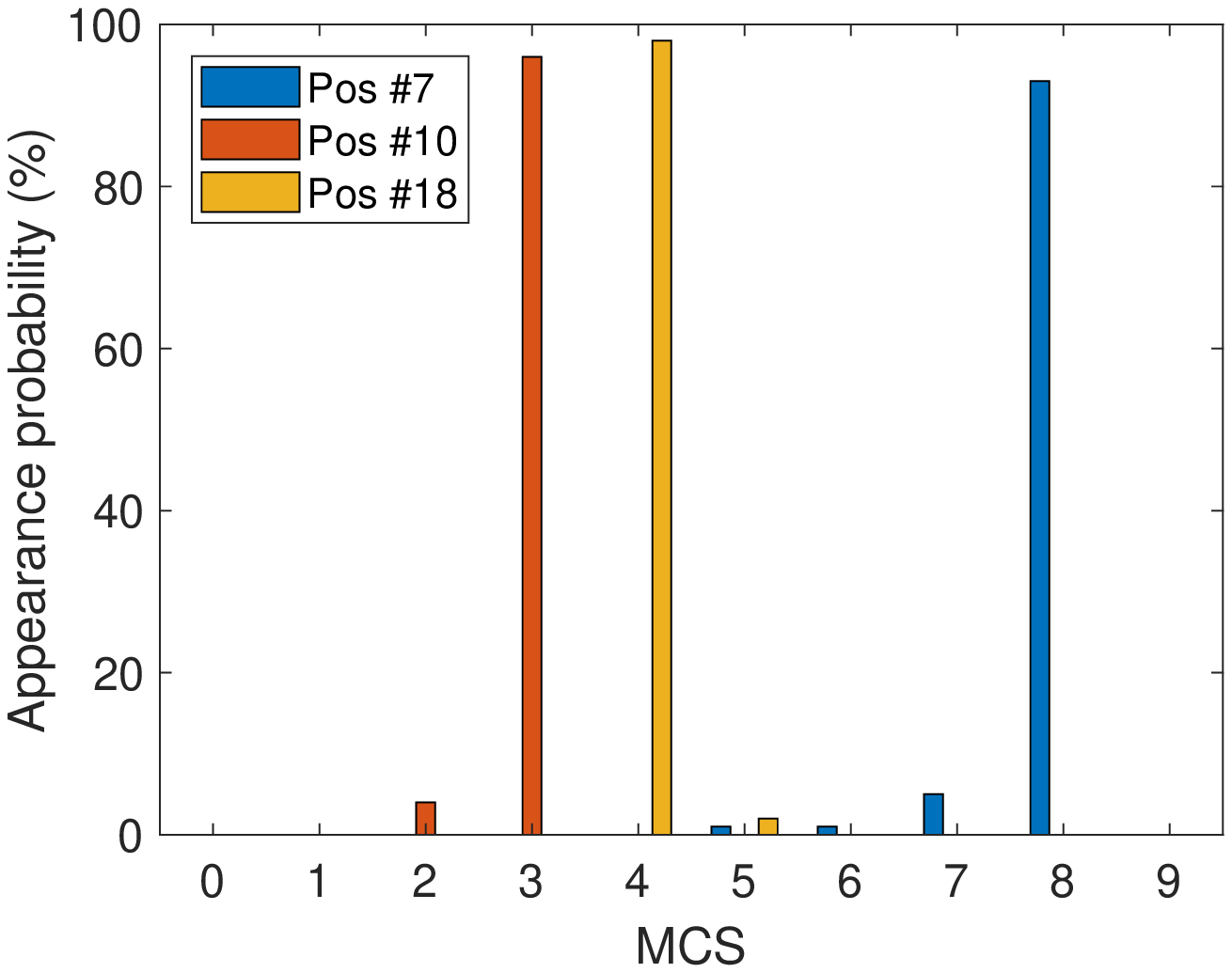}
                \caption{BW = 20 MHz, $P_{\text{TX}}$ = 10 dBm}
                \label{fig:MCS_20_10}
        \end{subfigure}
     \begin{subfigure}[b]{0.32\textwidth}
                \includegraphics[width=\textwidth]{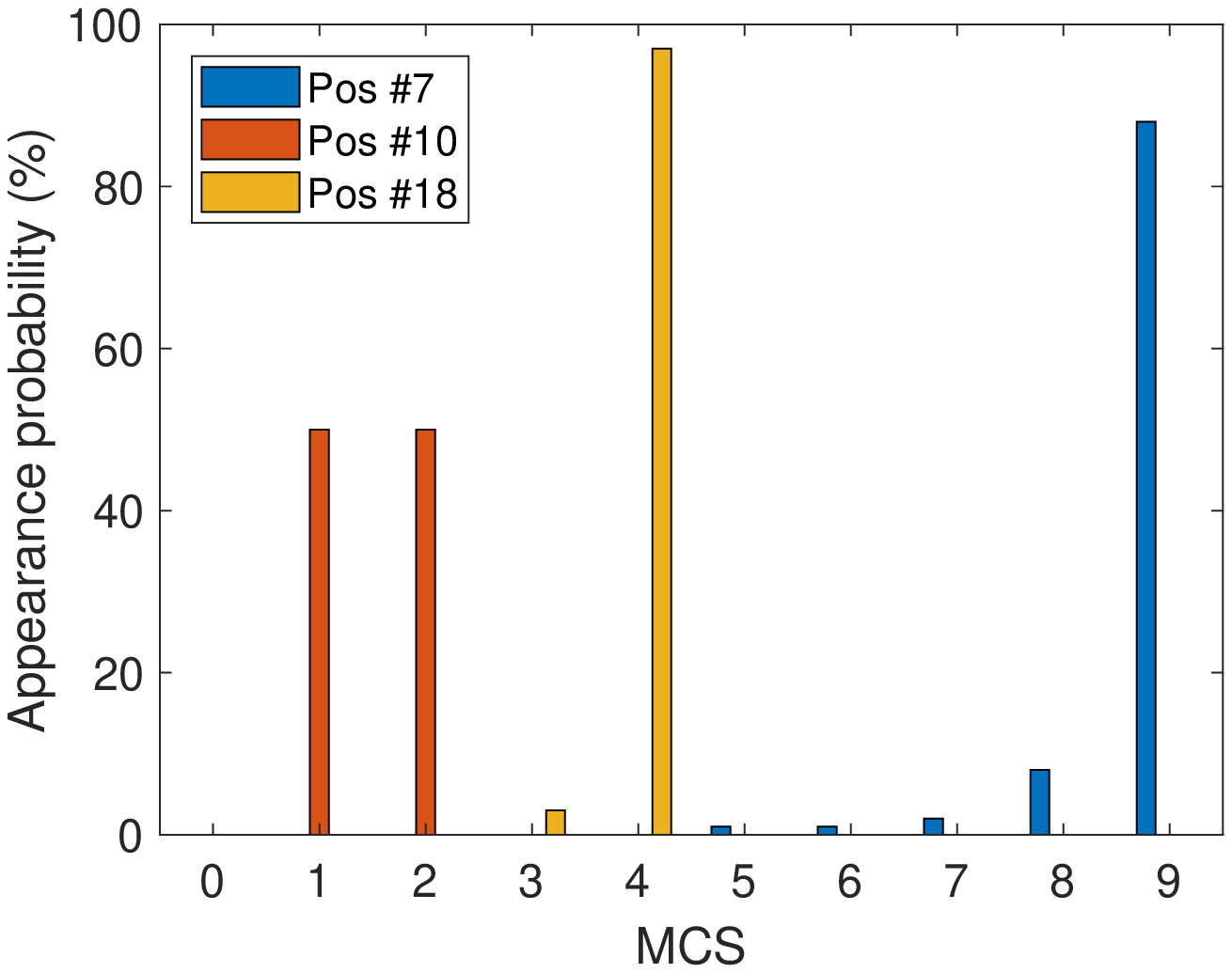}
                \caption{BW = 40 MHz, $P_{\text{TX}}$ = 10 dBm}
                \label{fig:MCS_40_10}
        \end{subfigure}   
      \begin{subfigure}[b]{0.32\textwidth}                \includegraphics[width=\textwidth]{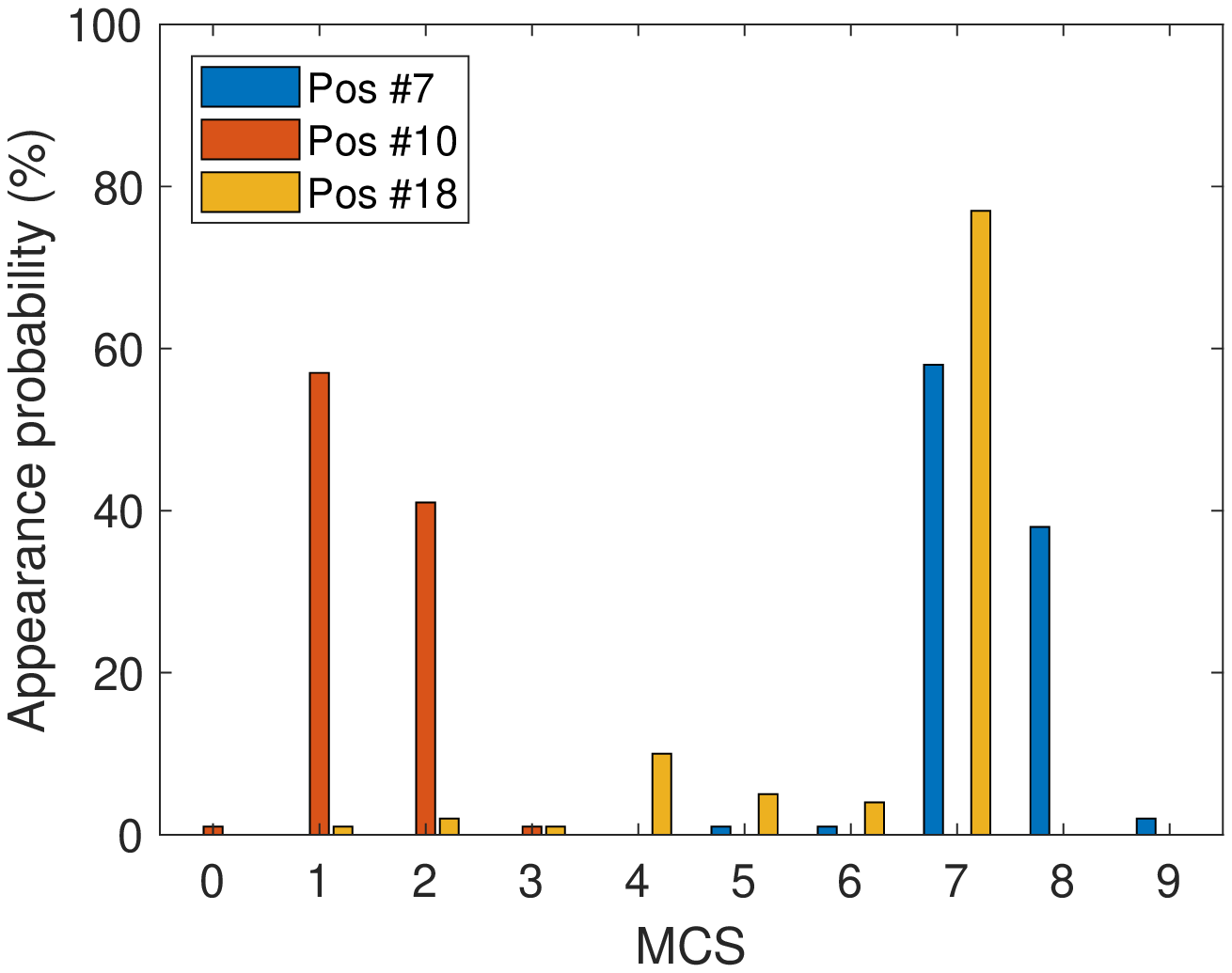}
                \caption{BW = 80 MHz, $P_{\text{TX}}$ = 10 dBm}
                \label{fig:MCS_80_10}
        \end{subfigure}
        
        \begin{subfigure}[b]{0.32\textwidth}
                \includegraphics[width=\textwidth]{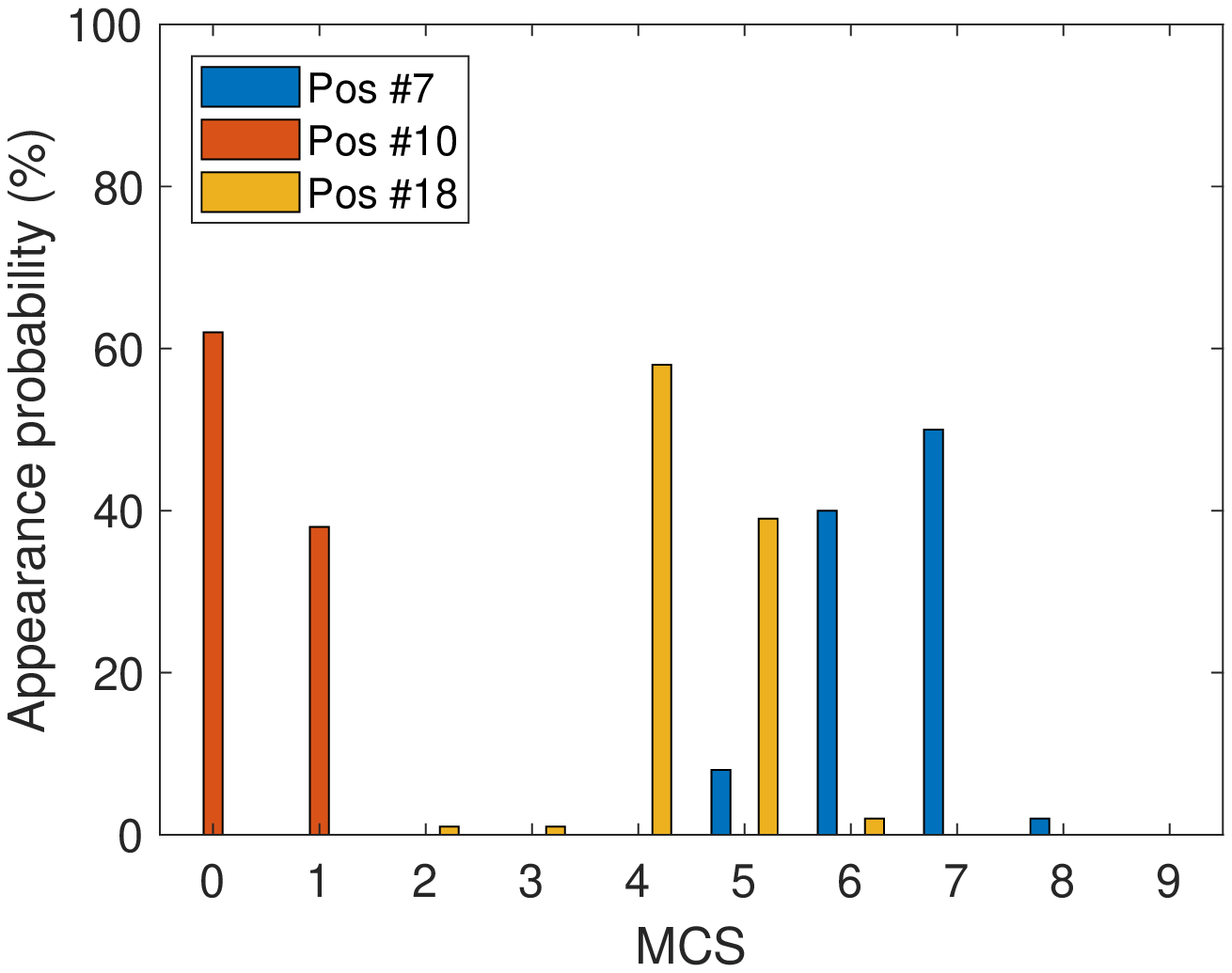}
                \caption{BW = 20 MHz, $P_{\text{TX}}$ = 4 dBm}
                \label{fig:MCS_20_4}
        \end{subfigure}
     \begin{subfigure}[b]{0.32\textwidth}
                \includegraphics[width=\textwidth]{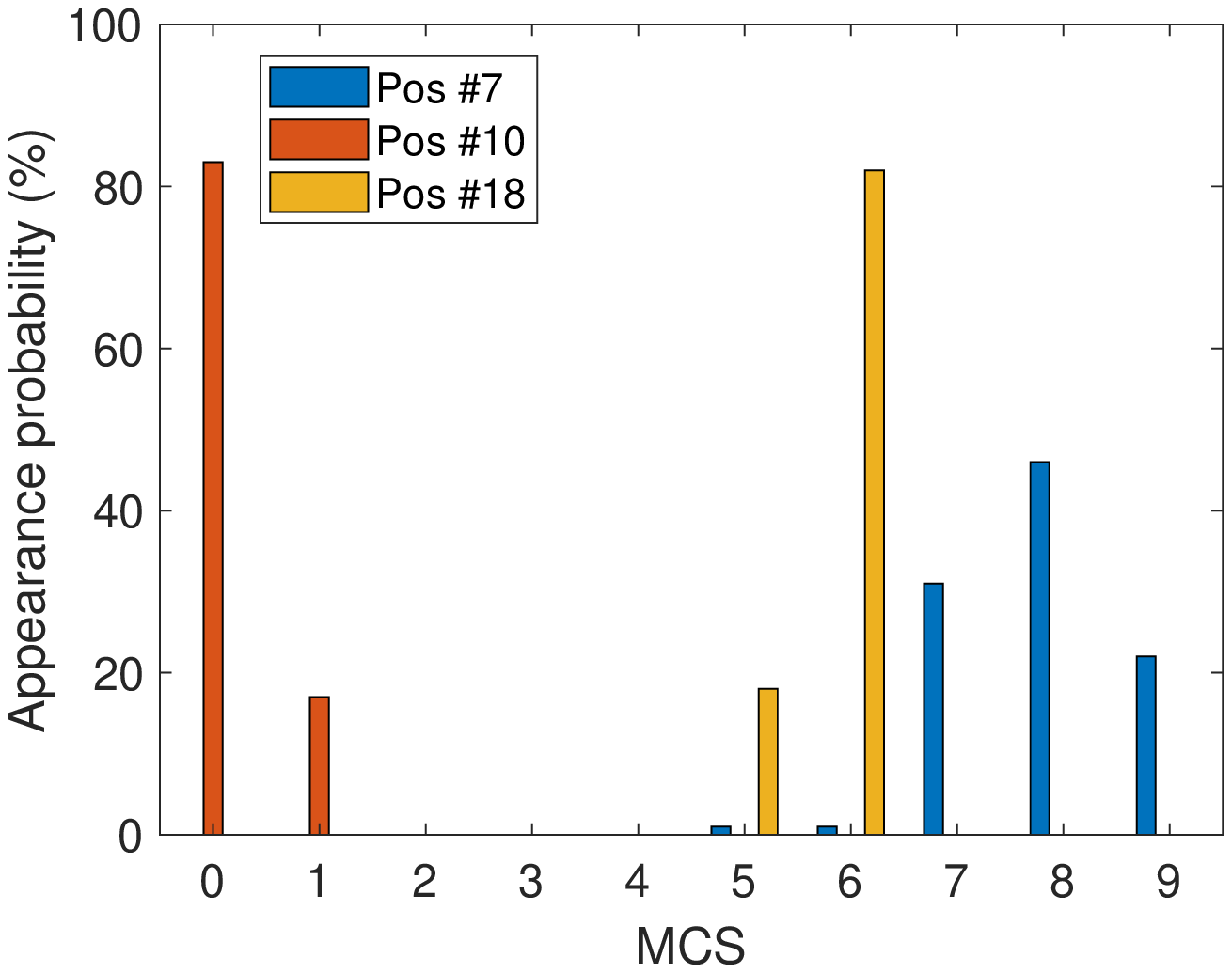}
                \caption{BW = 40 MHz, $P_{\text{TX}}$ = 4 dBm}
                \label{fig:MCS_40_4}
        \end{subfigure}   
      \begin{subfigure}[b]{0.32\textwidth}                \includegraphics[width=\textwidth]{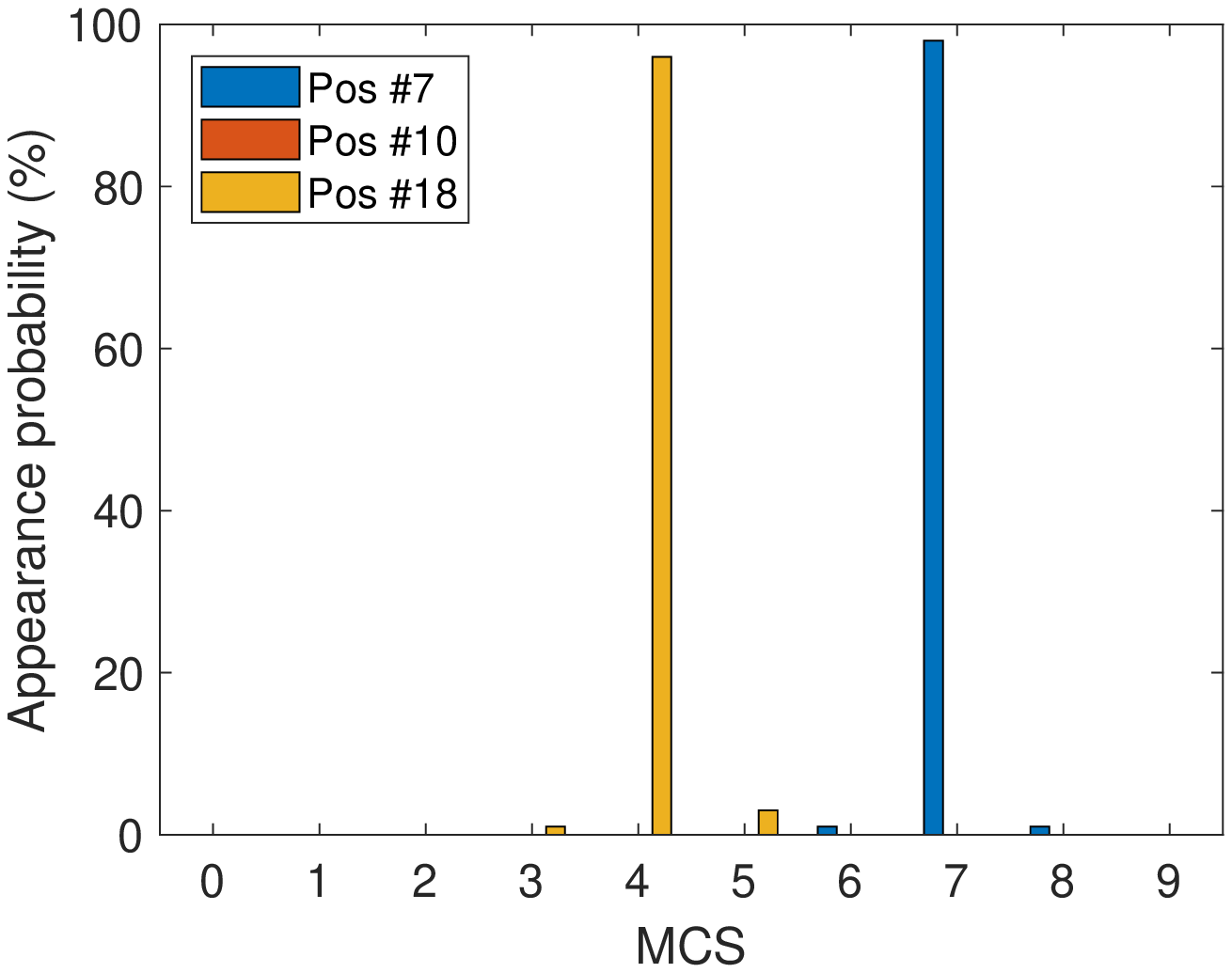}
                \caption{BW = 80 MHz, $P_{\text{TX}}$ = 4 dBm}
                \label{fig:MCS_80_4}
        \end{subfigure}
        \caption{MCS distribution in function of BW and $P_{\text{TX}}$ in Testbed \#2.}
        \label{fig:MCS}
\end{figure*}

Figure~\ref{fig:MCS} shows the appearance of each MCS (in \%) in every possible combination. The BW impact on the selected MCS follows a downward trend in most of the analyzed combinations (when not, it is due to a different number of spatial streams, as it will be studied in the next subsection).\footnote{Note the avoidance of MCS \#9 among tests ran with BW~=~ 20 MHz due to its unavailability in IEEE 802.11ac employed hardware.}

As expected, the AP tends to select greater MCSs when using higher $P_{\text{TX}}$ levels (specially in locations \#7 and \#10). As for location \#18, it does not completely follow this pattern, for example when using BW = 80 MHz (see Figures~\ref{fig:MCS_80_23}, \ref{fig:MCS_80_10}, and \ref{fig:MCS_80_4}). In this case, the highest MCSs are selected with $P_{\text{TX}}$ = 10 dBm, where only 1 spatial stream is used (unlike $P_{\text{TX}}$ = 23 dBm with 2 spatial streams).

%In this sense, this confirms that the MCS selection in IEEE 802.11ac (and therefore in IEEE 802.11ax) is no longer tied to the number of spatial streams, as it was in 802.11n \cite{gast2013802}. As already known, higher modulations pack more data into each transmission, but at the cost of requiring much higher signal-to-noise ratio (SNR). To boost SNR in poor-quality wireless channels, the AP makes use of its antennas to send redundant information using a single spatial stream, thus eluding spatial correlation \cite{halperin2010802}. Basically, when the AP reduces the number of spatial streams, the antennas are used to increase the \textit{diversity gain} at the cost of reducing the \textit{spatial multiplexing gain} \cite{zheng2003diversity}.

In this sense, this confirms that the MCS selection in IEEE 802.11ac (and therefore in IEEE 802.11ax) is no longer tied to the number of spatial streams, as it was in 802.11n \cite{gast2013802}. As already known, higher modulations pack more data into each transmission, but at the cost of requiring much higher signal-to-noise ratio (SNR). To boost SNR in poor-quality wireless channels, the AP makes use of its antennas to send redundant information using a single spatial stream, thus eluding spatial correlation \cite{halperin2010802}. Basically, when the AP reduces the number of spatial streams, the antennas are used to increase the \textit{diversity gain} at the cost of reducing the \textit{spatial multiplexing gain}.

\subsection{Spatial streams}
As noted in the previous section, the MCS selection process conducted by the AP cannot be detached from the number of employed spatial streams. This issue was analyzed in depth in Testbed \#1, where the receiver laptop stored the RSSI value, the number of spatial streams, and the MCS of each received packet.

Data obtained from the $N_{L}$ = 21 analyzed locations is aggregated and presented in Table~\ref{MCS_table}, offering the outcomes of the developed \textit{TMBmodel5GhzWIFI} MATLAB function:\footnote{The \textit{TMBmodel5GhzWIFI} MATLAB function is able to provide a vector of MCS probabilities in function of the AP-STA distance, the BW and the $P_{\text{TX}}$. It is available (together with the data sets obtained from described experimentation) in the following GitHub repository:\\ \url{https://github.com/wn-upf/TMBmodel5GhzWIFI}} a representation of the most selected MCS (i.e., the statistical \textit{mode}) and its associated number of spatial streams together with its appearance frequency (in \%) in function of the BW-$P_{\text{TX}}$ configuration of the AP and the collected RSSI in the receiver (here aggregated in 5-dB groups). Due to the non-homogeneous distribution of measurements along the receiver's RSSI range, there are some RSSI bands with fewer samples than others or even none, the latter case being represented with an empty cell.

Results show how MCSs with a single spatial stream are mainly below -72 dBm, while MCSs with two spatial streams are chosen in better channel conditions. However, the threshold between one and two spatial streams is not clearly defined, as shown with the two outliers located at RSSI bands between [-82, -78] dBm and [-77, -73] dBm. 

As expected, the index of the most selected MCS grows together with the RSSI in each analyzed AP configuration, covering all available modes except from the lowest ones using two spatial streams, which are less often used than those with a single stream and higher indexes. For a given RSSI band, however, there is no discernible tendency in the impact of BW and $P_{\text{TX}}$ on the selected MCS, even less in low RSSI values, where higher index diversity is observed.

Consequently, in the internal process of the AP to select the most appropriate MCS, the commercial character of the employed AP should be taken into consideration with respect to the conducted experimentation, as the actual MCS selection algorithm could contain some other rules and variables apart from BW,  $P_\text{TX}$, and RSSI.

\begin{table*}[]
\centering
\caption{MCS appearance frequency (in \%) in function of BW, $P_{\text{TX}}$, and RSSI. \newline (MCSs with 1 spatial stream are shown in \textit{blue}, whereas those using 2 spatial streams are shown in \textit{yellow}).}
\label{MCS_table}
\begin{tabular}{|c|c|c|c|c|c|c|c|c|c|}
\hline
                                                                                & \multicolumn{3}{c|}{\textbf{20 MHz}}                                                                                                                                                                                                                      & \multicolumn{3}{c|}{\textbf{40 MHz}}                                                                                                                                                                                                                                             & \multicolumn{3}{c|}{\textbf{80 MHz}}                                                                                                                                                                                                                       \\ \cline{2-10} 
\multirow{-2}{*}{\textbf{\begin{tabular}[c]{@{}c@{}}RSSI\\ (dBm)\end{tabular}}} & \textbf{4 dBm}                                                                    & \textbf{10 dBm}                                                                   & \textbf{23 dBm}                                                                   & \textbf{4 dBm}                                                                    & \textbf{10 dBm}                                                                                          & \textbf{23 dBm}                                                                   & \textbf{4 dBm}                                                                    & \textbf{10 dBm}                                                                   & \textbf{23 dBm}                                                                    \\ \hline
{[}-97, -93{]}                                                                  & \cellcolor[HTML]{DAE8FC}\begin{tabular}[c]{@{}c@{}}MCS 0\\ (82.42\%)\end{tabular} & \cellcolor[HTML]{DAE8FC}\begin{tabular}[c]{@{}c@{}}MCS 3\\ (54.57\%)\end{tabular} &                                                                                   &                                                                                   &                                                                                                          &                                                                                   &                                                                                   &                                                                                   &                                                                                    \\ \hline
{[}-92, -88{]}                                                                  & \cellcolor[HTML]{DAE8FC}\begin{tabular}[c]{@{}c@{}}MCS 2\\ (31.62\%)\end{tabular} & \cellcolor[HTML]{DAE8FC}\begin{tabular}[c]{@{}c@{}}MCS 3\\ (74.76\%)\end{tabular} &                                                                                   & \cellcolor[HTML]{DAE8FC}\begin{tabular}[c]{@{}c@{}}MCS 0\\ (51.24\%)\end{tabular} & \cellcolor[HTML]{DAE8FC}\begin{tabular}[c]{@{}c@{}}MCS 1\\ (51.41\%)\end{tabular}                        &                                                                                   & \cellcolor[HTML]{DAE8FC}\begin{tabular}[c]{@{}c@{}}MCS 1\\ (57.61\%)\end{tabular} & \cellcolor[HTML]{DAE8FC}\begin{tabular}[c]{@{}c@{}}MCS 0\\ (46.37\%)\end{tabular} & \cellcolor[HTML]{DAE8FC}\begin{tabular}[c]{@{}c@{}}MCS 1\\ (100.00\%)\end{tabular} \\ \hline
{[}-87, -83{]}                                                                  & \cellcolor[HTML]{DAE8FC}\begin{tabular}[c]{@{}c@{}}MCS 4\\ (33.10\%)\end{tabular} & \cellcolor[HTML]{DAE8FC}\begin{tabular}[c]{@{}c@{}}MCS 3\\ (55.00\%)\end{tabular} & \cellcolor[HTML]{DAE8FC}\begin{tabular}[c]{@{}c@{}}MCS 5\\ (42.86\%)\end{tabular} & \cellcolor[HTML]{DAE8FC}\begin{tabular}[c]{@{}c@{}}MCS 1\\ (69.45\%)\end{tabular} & \cellcolor[HTML]{DAE8FC}{\color[HTML]{333333} \begin{tabular}[c]{@{}c@{}}MCS 1\\ (34.75\%)\end{tabular}} & \cellcolor[HTML]{DAE8FC}\begin{tabular}[c]{@{}c@{}}MCS 3\\ (99.33\%)\end{tabular} & \cellcolor[HTML]{DAE8FC}\begin{tabular}[c]{@{}c@{}}MCS 1\\ (48.37\%)\end{tabular} & \cellcolor[HTML]{DAE8FC}\begin{tabular}[c]{@{}c@{}}MCS 2\\ (54.91\%)\end{tabular} & \cellcolor[HTML]{DAE8FC}\begin{tabular}[c]{@{}c@{}}MCS 1\\ (43.62\%)\end{tabular}  \\ \hline
{[}-82, -78{]}                                                                  & \cellcolor[HTML]{DAE8FC}\begin{tabular}[c]{@{}c@{}}MCS 5\\ (45.33\%)\end{tabular} & \cellcolor[HTML]{DAE8FC}\begin{tabular}[c]{@{}c@{}}MCS 6\\ (27.27\%)\end{tabular} & \cellcolor[HTML]{FFFC9E}\begin{tabular}[c]{@{}c@{}}MCS 3\\ (29.33\%)\end{tabular} & \cellcolor[HTML]{DAE8FC}\begin{tabular}[c]{@{}c@{}}MCS 4\\ (54.72\%)\end{tabular} & \cellcolor[HTML]{DAE8FC}\begin{tabular}[c]{@{}c@{}}MCS 3\\ (60.33\%)\end{tabular}                        & \cellcolor[HTML]{DAE8FC}\begin{tabular}[c]{@{}c@{}}MCS 5\\ (31.47\%)\end{tabular} & \cellcolor[HTML]{DAE8FC}\begin{tabular}[c]{@{}c@{}}MCS 3\\ (82.68\%)\end{tabular} & \cellcolor[HTML]{DAE8FC}\begin{tabular}[c]{@{}c@{}}MCS 4\\ (31.15\%)\end{tabular} & \cellcolor[HTML]{DAE8FC}\begin{tabular}[c]{@{}c@{}}MCS 3\\ (57.37\%)\end{tabular}  \\ \hline
{[}-77, -73{]}                                                                  & \cellcolor[HTML]{DAE8FC}\begin{tabular}[c]{@{}c@{}}MCS 4\\ (35.76\%)\end{tabular} & \cellcolor[HTML]{DAE8FC}\begin{tabular}[c]{@{}c@{}}MCS 5\\ (29.85\%)\end{tabular} & \cellcolor[HTML]{DAE8FC}\begin{tabular}[c]{@{}c@{}}MCS 5\\ (30.89\%)\end{tabular} & \cellcolor[HTML]{DAE8FC}\begin{tabular}[c]{@{}c@{}}MCS 4\\ (45.90\%)\end{tabular} & \cellcolor[HTML]{DAE8FC}\begin{tabular}[c]{@{}c@{}}MCS 6\\ (17.61\%)\end{tabular}                        & \cellcolor[HTML]{DAE8FC}\begin{tabular}[c]{@{}c@{}}MCS 5\\ (45.14\%)\end{tabular} & \cellcolor[HTML]{DAE8FC}\begin{tabular}[c]{@{}c@{}}MCS 4\\ (81.67\%)\end{tabular} & \cellcolor[HTML]{DAE8FC}\begin{tabular}[c]{@{}c@{}}MCS 6\\ (49.59\%)\end{tabular} & \cellcolor[HTML]{FFFC9E}\begin{tabular}[c]{@{}c@{}}MCS 4\\ (35.14\%)\end{tabular}  \\ \hline
{[}-72, -68{]}                                                                  & \cellcolor[HTML]{FFFC9E}\begin{tabular}[c]{@{}c@{}}MCS 7\\ (44.44\%)\end{tabular} & \cellcolor[HTML]{FFFC9E}\begin{tabular}[c]{@{}c@{}}MCS 6\\ (36.17\%)\end{tabular} & \cellcolor[HTML]{FFFC9E}\begin{tabular}[c]{@{}c@{}}MCS 7\\ (37.24\%)\end{tabular} & \cellcolor[HTML]{FFFC9E}\begin{tabular}[c]{@{}c@{}}MCS 4\\ (34.29\%)\end{tabular} & \cellcolor[HTML]{DAE8FC}\begin{tabular}[c]{@{}c@{}}MCS 7\\ (47.03\%)\end{tabular}                        & \cellcolor[HTML]{DAE8FC}\begin{tabular}[c]{@{}c@{}}MCS 6\\ (41.91\%)\end{tabular} & \cellcolor[HTML]{DAE8FC}\begin{tabular}[c]{@{}c@{}}MCS 7\\ (40.04\%)\end{tabular} & \cellcolor[HTML]{DAE8FC}\begin{tabular}[c]{@{}c@{}}MCS 7\\ (67.49\%)\end{tabular} & \cellcolor[HTML]{DAE8FC}\begin{tabular}[c]{@{}c@{}}MCS 8\\ (47.39\%)\end{tabular}  \\ \hline
{[}-67, -63{]}                                                                  & \cellcolor[HTML]{FFFC9E}\begin{tabular}[c]{@{}c@{}}MCS 8\\ (77.39\%)\end{tabular} & \cellcolor[HTML]{FFFC9E}\begin{tabular}[c]{@{}c@{}}MCS 6\\ (54.10\%)\end{tabular} & \cellcolor[HTML]{FFFC9E}\begin{tabular}[c]{@{}c@{}}MCS 5\\ (28.45\%)\end{tabular} & \cellcolor[HTML]{FFFC9E}\begin{tabular}[c]{@{}c@{}}MCS 8\\ (48.93\%)\end{tabular} & \cellcolor[HTML]{FFFC9E}\begin{tabular}[c]{@{}c@{}}MCS 7\\ (45.38\%)\end{tabular}                        & \cellcolor[HTML]{FFFC9E}\begin{tabular}[c]{@{}c@{}}MCS 4\\ (44.30\%)\end{tabular} & \cellcolor[HTML]{FFFC9E}\begin{tabular}[c]{@{}c@{}}MCS 7\\ (58.14\%)\end{tabular} & \cellcolor[HTML]{FFFC9E}\begin{tabular}[c]{@{}c@{}}MCS 4\\ (42.02\%)\end{tabular} & \cellcolor[HTML]{FFFC9E}\begin{tabular}[c]{@{}c@{}}MCS 5\\ (61.79\%)\end{tabular}  \\ \hline
{[}-62, -58{]}                                                                  & \cellcolor[HTML]{FFFC9E}\begin{tabular}[c]{@{}c@{}}MCS 8\\ (60.70\%)\end{tabular} & \cellcolor[HTML]{FFFC9E}\begin{tabular}[c]{@{}c@{}}MCS 8\\ (86.00\%)\end{tabular} & \cellcolor[HTML]{FFFC9E}\begin{tabular}[c]{@{}c@{}}MCS 7\\ (71.37\%)\end{tabular} & \cellcolor[HTML]{FFFC9E}\begin{tabular}[c]{@{}c@{}}MCS 9\\ (51.95\%)\end{tabular} & \cellcolor[HTML]{FFFC9E}\begin{tabular}[c]{@{}c@{}}MCS 9\\ (65.36\%)\end{tabular}                        & \cellcolor[HTML]{FFFC9E}\begin{tabular}[c]{@{}c@{}}MCS 9\\ (55.56\%)\end{tabular} & \cellcolor[HTML]{FFFC9E}\begin{tabular}[c]{@{}c@{}}MCS 7\\ (62.26\%)\end{tabular} & \cellcolor[HTML]{FFFC9E}\begin{tabular}[c]{@{}c@{}}MCS 9\\ (63.79\%)\end{tabular} & \cellcolor[HTML]{FFFC9E}\begin{tabular}[c]{@{}c@{}}MCS 8\\ (45.01\%)\end{tabular}  \\ \hline
{[}-57, -53{]}                                                                  & \cellcolor[HTML]{FFFC9E}\begin{tabular}[c]{@{}c@{}}MCS 8\\ (50.33\%)\end{tabular} & \cellcolor[HTML]{FFFC9E}\begin{tabular}[c]{@{}c@{}}MCS 8\\ (99.13\%)\end{tabular} & \cellcolor[HTML]{FFFC9E}\begin{tabular}[c]{@{}c@{}}MCS 8\\ (66.46\%)\end{tabular} & \cellcolor[HTML]{FFFC9E}\begin{tabular}[c]{@{}c@{}}MCS 8\\ (60.79\%)\end{tabular} & \cellcolor[HTML]{FFFC9E}\begin{tabular}[c]{@{}c@{}}MCS 9\\ (93.40\%)\end{tabular}                        & \cellcolor[HTML]{FFFC9E}\begin{tabular}[c]{@{}c@{}}MCS 8\\ (52.76\%)\end{tabular} & \cellcolor[HTML]{FFFC9E}\begin{tabular}[c]{@{}c@{}}MCS 7\\ (68.81\%)\end{tabular} & \cellcolor[HTML]{FFFC9E}\begin{tabular}[c]{@{}c@{}}MCS 9\\ (74.51\%)\end{tabular} &                                                                                    \\ \hline
{[}-52, -48{]}                                                                  & \cellcolor[HTML]{FFFC9E}\begin{tabular}[c]{@{}c@{}}MCS 8\\ (97.92\%)\end{tabular} & \cellcolor[HTML]{FFFC9E}\begin{tabular}[c]{@{}c@{}}MCS 8\\ (95.97\%)\end{tabular} & \cellcolor[HTML]{FFFC9E}\begin{tabular}[c]{@{}c@{}}MCS 8\\ (99.12\%)\end{tabular} & \cellcolor[HTML]{FFFC9E}\begin{tabular}[c]{@{}c@{}}MCS 9\\ (53.30\%)\end{tabular} & \cellcolor[HTML]{FFFC9E}\begin{tabular}[c]{@{}c@{}}MCS 9\\ (93.35\%)\end{tabular}                        & \cellcolor[HTML]{FFFC9E}\begin{tabular}[c]{@{}c@{}}MCS 9\\ (95.55\%)\end{tabular} & \cellcolor[HTML]{FFFC9E}\begin{tabular}[c]{@{}c@{}}MCS 9\\ (94.86\%)\end{tabular} & \cellcolor[HTML]{FFFC9E}\begin{tabular}[c]{@{}c@{}}MCS 7\\ (57.39\%)\end{tabular} & \cellcolor[HTML]{FFFC9E}\begin{tabular}[c]{@{}c@{}}MCS 9\\ (96.58\%)\end{tabular}  \\ \hline
{[}-47, -43{]}                                                                  & \cellcolor[HTML]{FFFC9E}\begin{tabular}[c]{@{}c@{}}MCS 8\\ (98.51\%)\end{tabular} & \cellcolor[HTML]{FFFC9E}\begin{tabular}[c]{@{}c@{}}MCS 8\\ (97.89\%)\end{tabular} & \cellcolor[HTML]{FFFC9E}\begin{tabular}[c]{@{}c@{}}MCS 8\\ (99.07\%)\end{tabular} & \cellcolor[HTML]{FFFC9E}\begin{tabular}[c]{@{}c@{}}MCS 9\\ (95.37\%)\end{tabular} & \cellcolor[HTML]{FFFC9E}\begin{tabular}[c]{@{}c@{}}MCS 9\\ (52.91\%)\end{tabular}                        & \cellcolor[HTML]{FFFC9E}\begin{tabular}[c]{@{}c@{}}MCS 9\\ (95.82\%)\end{tabular} & \cellcolor[HTML]{FFFC9E}\begin{tabular}[c]{@{}c@{}}MCS 9\\ (97.69\%)\end{tabular} & \cellcolor[HTML]{FFFC9E}\begin{tabular}[c]{@{}c@{}}MCS 9\\ (93.71\%)\end{tabular} & \cellcolor[HTML]{FFFC9E}\begin{tabular}[c]{@{}c@{}}MCS 9\\ (91.60\%)\end{tabular}  \\ \hline
{[}-42, -38{]}                                                                  &                                                                                   & \cellcolor[HTML]{FFFC9E}\begin{tabular}[c]{@{}c@{}}MCS 8\\ (97.25\%)\end{tabular} & \cellcolor[HTML]{FFFC9E}\begin{tabular}[c]{@{}c@{}}MCS 8\\ (96.00\%)\end{tabular} & \cellcolor[HTML]{FFFC9E}\begin{tabular}[c]{@{}c@{}}MCS 9\\ (99.58\%)\end{tabular} & \cellcolor[HTML]{FFFC9E}\begin{tabular}[c]{@{}c@{}}MCS 9\\ (98.58\%)\end{tabular}                        & \cellcolor[HTML]{FFFC9E}\begin{tabular}[c]{@{}c@{}}MCS 9\\ (85.96\%)\end{tabular} &                                                                                   & \cellcolor[HTML]{FFFC9E}\begin{tabular}[c]{@{}c@{}}MCS 9\\ (97.35\%)\end{tabular} & \cellcolor[HTML]{FFFC9E}\begin{tabular}[c]{@{}c@{}}MCS 9\\ (87.26\%)\end{tabular}  \\ \hline
{[}-37, -33{]}                                                                  &                                                                                   &                                                                                   & \cellcolor[HTML]{FFFC9E}\begin{tabular}[c]{@{}c@{}}MCS 8\\ (99.55\%)\end{tabular} &                                                                                   & \cellcolor[HTML]{FFFC9E}\begin{tabular}[c]{@{}c@{}}MCS 9\\ (99.12\%)\end{tabular}                        & \cellcolor[HTML]{FFFC9E}\begin{tabular}[c]{@{}c@{}}MCS 9\\ (90.21\%)\end{tabular} &                                                                                   &                                                                                   &                                                                                    \\ \hline
{[}-32, -28{]}                                                                  &                                                                                   &                                                                                   & \cellcolor[HTML]{FFFC9E}\begin{tabular}[c]{@{}c@{}}MCS 8\\ (64.42\%)\end{tabular} &                                                                                   &                                                                                                          & \cellcolor[HTML]{FFFC9E}\begin{tabular}[c]{@{}c@{}}MCS 9\\ (97.21\%)\end{tabular} &                                                                                   &                                                                                   & \cellcolor[HTML]{FFFC9E}\begin{tabular}[c]{@{}c@{}}MCS 9\\ (98.16\%)\end{tabular}  \\ \hline
{[}-27, -23{]}                                                                  &                                                                                   &                                                                                   & \cellcolor[HTML]{FFFC9E}\begin{tabular}[c]{@{}c@{}}MCS 8\\ (97.82\%)\end{tabular} &                                                                                   &                                                                                                          &                                                                                   &                                                                                   &                                                                                   &                                                                                    \\ \hline
\end{tabular}
\end{table*}

\section{Conclusions}
\label{conclusions}
IEEE 802.11ax is an important step forward for WiFi, bringing many features and improvements to support multi-user and high-throughput application requirements. However, the inherent signal propagation characteristics of the indoor scenarios in which these networks are planned to be deployed complicate prior planning and physical network dimensioning.

Accurate indoor path loss models could help researchers to better understand IEEE 802.11ax PHY and MAC layers, which would lead to the development of better tools to simulate their behaviour, detect inefficiencies, and propose novel technological improvements. Similarly, operators and installers could also benefit from these advanced tools prior to network planning and deployment operations.

This article provides the new TMB path loss model for 5 GHz indoor IEEE 802.11ac/ax scenarios. Designed as a \textit{continuous} model, the TMB model does not require from previous computation of traversed obstacles to provide the path loss value for a given AP-STA distance, in contrast to existing \textit{location-specific} models. In fact, the model suitability in indoor environments has been proved by means of extensive experimentation in a typical office floor configuration with multiple room partition walls.

%This article provides the new TMB path loss model for IEEE 802.11ac/ax that reduces the number of \textit{in situ} measurements required to characterize an indoor scenario in comparison to the already available models from the TGax. \textcolor{red}{[No em queda clar que es vol dir aquí. Es redueix el número de mesures in-situ??? La idea es fer-lo servir sense fer cap mesura...]} Its suitability has been proved in real testbeds, achieving better accuracy with respect to measured values than traditional \textit{continuous} \textcolor{red}{[estem fent servir les mateixes dades del training per l'avaluació???? :P Normal que vagi millor que altres models... jo no ho diria així. Potser fer més emfasis que el TMB model es representatiu d'escenaris com el nostre]}, and even \textit{location-specific} TGax models.

A comprehensive study on the empirical relationship between RSSI and MCS has shown how low BW and high $P_{\text{TX}}$ levels configured in the AP lead to larger modulation indexes. As for the number of spatial streams, they grow together with the RSSI in all studied cases. Both facts are considered in the  developed \textit{TMBmodel5GhzWIFI} function, which returns the MCS and number of spatial streams distribution according to the AP-STA distance and the BW-$P_{\text{TX}}$ configuration. 

Multi-floor environments and technical implications of upcoming commercial IEEE 802.11ax devices (for instance, the use of MU-MIMO technology) stand as the two main issues to be integrated into the TMB model in the near future. Additionally, the current study paves the way of an in-depth analysis of the channel occupancy rate (COR) in a WLAN in function of the receiver's location.

%Improvement of the proposed path loss model stands as the main challenge of the near future, taking into consideration both the effect of building floors on the received signal and the implications of upcoming commercial IEEE 802.11ax devices.  \textcolor{red}{[Jo aquí diria veure com funciona MU-MIMO, i si el pathloss quan fas MU beamforming es calcula de manera similar.]}


%%%%%%%%%%%%%%%%%%%%%%%%%%%%%%%%%%%%%%%%%%%%%%%%%%%%%%%%%%%%%%%%%%%%%%%%%%%%%%%%

%%%%%%%%%%%%%%%%%%%%%%%%%%%%%%%%%%%%%%%%%%%%%%%%%%%%%%%%%%%%%%%%%%%%%%%%%%%%%%%%

%%%%%%%%%%%%%%%%%%%%%%%%%%%%%%%%%%%%%%%%%%%%%%%%%%%%%%%%%%%%%%%%%%%%%%%%%%%%%%%%

\section*{Acknowledgement}

This work was partially supported by the Cisco University Research Program fund (Project CG No. 890107, Towards Deterministic Channel Access in High-Density WLANs), a corporate advised fund of Silicon Valley Community Foundation. It also received funding from the Catalan government through projects SGR-2017-1188 and SGR-2017-1739, and from the Spanish government under the project TEC2016-79510-P.

%%%%%%%%%%%%%%%%%%%%%%%%%%%%%%%%%%%%%%%%%%%%%%%%%%%%%%%%%%%%%%%%%%%%%%%%%%%%%%%%

\bibliography{Bib}

\begin{thebibliography}{10}

\bibitem{tgax2018}
{IEEE 802.11 Task Group AX}, ``{Status of Project IEEE 802.11ax High Efficiency
  WLAN (HEW)}.'' \url{http://www.ieee802.org/11/Reports/tgax_update.htm}.
\newblock {Accessed: 2018-11-29}.

\bibitem{bellalta2016ieee}
B.~Bellalta, ``{IEEE 802.11 ax: High-efficiency WLANs},'' {\em IEEE Wireless
  Communications}, vol.~23, no.~1, pp.~38--46, 2016.

\bibitem{qualcomm2018trans}
{Qualcomm}, ``{802.11ax: Transforming Wi-Fi to bring unprecedented capacity and
  efficiency}.''
  \url{https://www.qualcomm.com/media/documents/files/802-11ax-transforming-wi-fi-to-bring-unprecedented-capacity-efficiency.pdf}.
\newblock {Accessed: 2018-11-29}.

\bibitem{lott2001multi}
M.~Lott and I.~Forkel, ``A multi-wall-and-floor model for indoor radio
  propagation,'' in {\em Vehicular Technology Conference, 2001. VTC 2001
  Spring. IEEE VTS 53rd}, vol.~1, pp.~464--468, IEEE, 2001.

\bibitem{keenan1990radio}
J.~Keenan and A.~Motley, ``Radio coverage in buildings,'' {\em British telecom
  technology Journal}, vol.~8, no.~1, pp.~19--24, 1990.

\bibitem{solahuddin2011indoor}
Y.~Solahuddin and R.~Mardeni, ``{Indoor empirical path loss prediction model
  for 2.4 GHz 802.11n network},'' in {\em Control System, Computing and
  Engineering (ICCSCE), 2011 IEEE International Conference on}, pp.~12--17,
  IEEE, 2011.

\bibitem{xu2007indoor}
D.~Xu, J.~Zhang, X.~Gao, P.~Zhang, and Y.~Wu, ``{Indoor office propagation
  measurements and path loss models at 5.25 GHz},'' in {\em Vehicular
  Technology Conference, 2007. VTC-2007 Fall. 2007 IEEE 66th}, pp.~844--848,
  IEEE, 2007.

\bibitem{tgax2018channel}
{IEEE 802.11 Task Group AX}, ``{IEEE 802.11ax Channel Model Document}.''
  \url{https://mentor.ieee.org/802.11/dcn/14/11-14-0882-04-00ax-tgax-channel-model-document.docx}.
\newblock {Accessed: 2018-11-29}.

\bibitem{tgax2018simulation}
{IEEE 802.11 Task Group AX}, ``{TGax Simulation Scenarios}.''
  \url{https://mentor.ieee.org/802.11/dcn/14/11-14-0980-14-00ax-simulationscenarios.docx}.
\newblock {Accessed: 2018-11-29}.

\bibitem{afaqui2016ieee}
M.~S. Afaqui, E.~Villegas, and E.~Aguilera, ``{IEEE 802.11ax: Challenges and
  requirements for future high efficiency WiFi},'' {\em IEEE Wireless
  Communications}, vol.~99, pp.~2--9, 2016.

\bibitem{wilhelmi2018potential}
F.~Wilhelmi, S.~Barrachina-Mu{\~n}oz, C.~Cano, B.~Bellalta, A.~Jonsson, and
  G.~Neu, ``{Potential and Pitfalls of Multi-Armed Bandits for Decentralized
  Spatial Reuse in WLANs},'' {\em arXiv preprint arXiv:1805.11083}, 2018.

\bibitem{itu2012p}
ITU-R, ``{Recommendation P.1238-7: Propagation data and prediction methods for
  the planning of indoor radio communication systems and radio local area
  networks in the frequency range 900 MHz to 100 GHz},'' {\em P Series.
  Radiowave propagation}, 2012.

\bibitem{gast2013802}
M.~S. Gast, {\em {802.11 ac: a survival guide: Wi-Fi at gigabit and beyond}}.
\newblock O'Reilly Media, Inc., 2013.

\bibitem{halperin2010802}
D.~Halperin, W.~Hu, A.~Sheth, and D.~Wetherall, ``802.11 with multiple antennas
  for dummies,'' {\em ACM SIGCOMM Computer Communication Review}, vol.~40,
  no.~1, pp.~19--25, 2010.

\end{thebibliography}
\bibliographystyle{ieeetr}

\end{document}